\documentclass[12pt]{article}
\usepackage[body={16cm,25cm}]{geometry}
\usepackage{amssymb}
\usepackage[margin=10pt,font=small,labelfont=bf,
labelsep=endash]{caption}

\renewcommand{\theequation}{\thesection.\arabic{equation}}

\newcommand{\be}{\begin{equation}}
\newcommand{\ee}{\end{equation}}
\newcommand{\bea}{\begin{eqnarray}}
\newcommand{\eea}{\end{eqnarray}}

\def\ds{\displaystyle}
\def\nsection#1{\setcounter{equation}{0}\section{#1}}

\def\STR{star-triangle relation}
\def\YBE{Yang-Baxter equation}
\def\DQSGM{discrete quantum Sine-Gordon model}
\def\e{\mbox{e}}
\def\ib{\,\mbox{i}\,}

\def\w{\omega}

\def\d{\delta}
\def\a{\alpha}
\def\b{\beta}
\def\l{\lambda}
\def\q{\mathsf q}

\def\HT{\ds{\widetilde H}}

\def\fla#1{(\ref{#1})}
\newcommand{\Wbar}{\overline{W}}
\def\lra{\longrightarrow}
\def\llra{\longleftrightarrow}
\begin{document}\renewcommand{\arraystretch}{1.5}
\newcounter{storeeqn}
\newenvironment{numlet}{
\addtocounter{equation}{1}
\setcounter{storeeqn}{\value{equation}}
\setcounter{equation}{0}
\renewcommand{\theequation}{\thesection.\arabic{storeeqn}\alph{equation}}}{
\setcounter{equation}{\value{storeeqn}}
\renewcommand{\theequation}{\thesection.\arabic{equation}}}
%\renewcommand{\baselinestretch}{2}

%%%%%%%%%%%%%NUMERATION IN APPENDICES AND SUB-APPENDICES%%%%%%%%%%%%%%%%%
\newcounter{app}
\newcounter{sapp}[app]
\def\theapp{\Alph{app}}
\newcommand{\app}[1]{
\refstepcounter{app}{\vspace{7mm}
\noindent\Large\bf Appendix
\theapp.
 \ #1 \par \vspace{5mm}}
\setcounter{equation}{0}
\def\theequation{\Alph{app}.\arabic{equation}}}
\def\thesapp{\Alph{app}.\arabic{sapp}}
\newcommand{\sapp}[1]{\par \refstepcounter{sapp}{\noindent\large\bf \thesapp
\ #1 \par \vspace{3mm}}
\def\theequation{\Alph{app}.\arabic{equation}}}
%%%%%%%%%%%%%%%%%%%%%%%%%%%%%%%%%%%%%%%%%%%%%%%%%%%%%%%%%%%%%%%%%%%

\renewcommand{\author}[1]{\large\rm #1\\ \bigskip}
\newcommand{\address}[1]{{\normalsize\it #1\\}\bigskip}
\renewcommand{\title}[1]{\bigskip\bigskip\Large\bf #1\bigskip\bigskip\\}

%  #] Preamble:
%\begin{titlepage}
\vglue 2cm

\begin{center}
%  #[ Title and author:
\title{Chiral Potts model and the discrete \\
Sine-Gordon model at roots of unity}

\vspace{1cm}

\author{        Vladimir V. Bazhanov\footnote[1]{email:
                {\tt Vladimir.Bazhanov(at)anu.edu.au}}}

\address{Department of Theoretical Physics,\\
         Research School of Physical Sciences and Engineering,\\
    Australian National University, Canberra, ACT 0200, Australia.}

\end{center}

\date{\ }
%\maketitle
%  #] Title and author:
\vspace{3cm}
%  #[ Abstract:
\begin{abstract}
The discrete quantum 
Sine-Gordon model at roots of unity remarkably combines a classical 
integrable system with an integrable quantum spin system, 
whose parameters obey classical equations of motion.
We show that the fundamental $R$-matrix of the model (which satisfies
a difference property \YBE ) naturally splits into a product
of a singular ``classical'' part and a finite dimensional quantum part.
The classical part of the $R$-matrix itself satisfies the quantum \YBE ,
and therefore can be factored out producing, however, a certain ``twist''
of the quantum part. We show that the resulting equation exactly coincides
with the star-triangle relation of the $N$-state chiral Potts model.
The associated spin model on the whole lattice is, in fact, 
more general than the chiral Potts and reduces to the latter only for
the simplest (constant) classical background. 
In a general case the model is inhomogeneous: its Boltzmann 
weights are determined by non-trivial background solutions of 
the equations of motion of the classical discrete sine-Gordon model.  

\end{abstract}
%\vspace{1cm}
%PACS numbers: 71,75, 02.
%\vspace{1cm}
%Running head: Lattice Sine-Gordon and Chiral Potts models.
%\end{titlepage}
\thispagestyle{empty}
%  #] Abstract:

%  #[ Introduction:
\newlength{\mathin}
\clearpage
\setcounter{page}{2}
\nsection {Introduction}
\def\r{}

The fundamental works 
of Professor Akihiro Tsuchiya \cite{Tsu86,Tsu89}  made an outstanding
contribution to the theory of integrable quantum systems. 
Here we present some new developments in this field. 
The discovery of the chiral Potts model originated in
\cite{HKdN,vG85,AuY87} 
and finalized in \cite{BPA88}  brought
a lot of new interesting
problems. First the Boltzmann weights of the model require high genus 
algebraic functions for their uniformization. The second (related) problem is 
that there is no difference property in the model. The  uniformization
problem has   been solved in \cite{CPH}
 using classical results 
on the rotation of a rigid body \cite{Kow}.
  However, even with this uniformization
we still do not know how to apply various methods which 
worked perfectly 
for all other two-dimensional solvable models, basically because there 
is  no a 
difference property. In particular, it took 16 years until the 
intriguingly simple  conjecture of \cite{Albertini:1988ux} 
for the local order parameter
\be
<\w^{\a j}>=(k^2)^{\Delta_j},\quad\Delta_j={j(N-j)\over2N^2},\quad
j=1,\ldots ,N-1.\label{magnetization}
\ee
has finally been proven \cite{Baxter:2005jt} by a generalization of
the method of \cite{JMN93}. 
Here $N$, $N\ge2$, is the number of local spin states 
and $k$ is a modulus of the the algebraic curve appearing in the model, it also
plays the role of the temperature like variable: the model is critical for 
$k=0$, it reduces than to the Fateev-Zamolodchikov model \cite{FZ82}.
For $N=2$ the chiral Potts  model coincides with the ordinary Ising model and 
formula \fla{magnetization} reduces  to the famous Onsager-Yang result 
\cite{O49,Y52} for the spontaneous magnetization of the Ising model.

In view of the above discussion the chiral Potts model may seem standing 
very far away from the ``conventional'' solvable model with the difference 
property. However, this is not so. It was shown \cite{BS90}
that the chiral Potts model is remarkably connected to the six-vertex model:
its R-matrix satisfies the \YBE s with two cyclic L-operators related to the 
R-matrix of the six-vertex model.  Further, it turned out that the last
connection can be extended  by replacing the six-vertex model
(with two states per edge) by an  the  $n$-state
model \cite{Cherednik:1980ey}  associated with the  $U_q({sl(n)})$ algebra.
This resulted in  
the ``$sl(n)$-generalized chiral Potts model" \cite{BKMS}, which is  a
two-dimensional model with spins that each take $N^{n-1}$ values.

Note, also that the chiral Potts model has very deep relations with the theory
of cyclic representations of quantum groups at roots of unity
\cite{DCK90,Date:1990bs}.  
In particular, the 
R-matrix of the model 
can be interpreted \cite{BKMS,Date:1990bs} 
as an intertwiner for two minimal cyclic representations
of the $U_q(\widehat{sl(n)})$ algebra. 

Next, in \cite{BB92} it was shown 
 that the $sl(n)$-generalized chiral Potts model 
can be interpreted as a model on a three-dimen\-sional simple cubic lattice  
consisting of $n$ square-lattice layers. At each site there is an $N$-valued 
spin. And when $N=2$ 
this three-dimensional model reduces 
(to within a minor modification of the boundary
conditions) to  the Zamolodchikov model \cite{Z80,Z81}. For $N>2$ it gives 
a new solvable interaction-round-a-cube  model in three dimensions \cite{BB92}.
The Boltzmann weights of this  model satisfy \cite{B83,KMS93} 
the tetrahedron relations which ensure its integrability. However,
it turns that the integrability of the model can be proved bypassing
the tetrahedron relations. It is based on a much simpler ``restricted 
star-triangle relation'' \cite{BB93}, which is a particular case of the
star-triangle relation \cite{BPA88,AYP89} 
for the original chiral Potts model. 
 
In this paper we discuss yet another remarkable connection of the
chiral Potts model. This time we relate it with the quantum discrete  
sine-Gordon (QDSG) model \cite{FV:1993,BKP,Faddeev:1994}. 
In \cite{BBR} it was
shown that for rational values of the coupling constant the QDSG-model
can be viewed as some special two-dimensional lattice model of statistical
mechanics where (integer) spins take $N\ge2$ values and where the Boltzmann 
weights are determined by solutions of the equation of motion of the {\em
  classical} discrete sine-Gordon model (the
situation here is very similar to quantum field theory in a 
classical ``background'' field).   
Here we show that in the simplest case of the constant background the
above lattice model exactly coincides with the chiral Potts model. 

The organization of the paper is as follows. In Section 2 we review
the formulation and basic properties of the classical and quantum
discrete sine-Gordon models. Associated solutions of the Yang-Baxter
equation are considered in Section 3. In Section 4 we review the basic
definitions of the chiral Potts model and Section 5 we relate this
model to the QDSG model.

\nsection{The quantum discrete Sine-Gordon model.}
\subsection{Formulation of the model.}
The Sine-Gordon equation for a
scalar function $\phi(x,t)$ in $1+1$ dimensions
\be
-\partial^2_t\phi + \partial^2_x\phi = m^2\sin\phi  \label{SGE}
\ee
is the most famous example of a system integrable on both the  
classical and quantum levels. Despite of an extensive literature
devoted to this model (see \cite{ZZ},\cite{FT} for the references), 
it still continues
to reveal its new features. 
In this paper we shall consider  integrable generalizations
\cite{FV:1993,BKP,Faddeev:1994} of this 
model (classical and quantum) for the case when both the space and time
variables are discrete. 

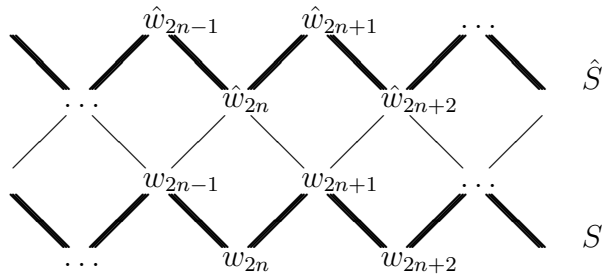
\begin{figure}[hbt]\begin{center}
\begin{picture}(230,120)
\thicklines
\multiput(40,20)(60,0){3}{\line(1,1){20}}
\multiput(10,40)(60,0){4}{\line(1,-1){20}}
\multiput(40,80)(60,0){3}{\line(1,1){20}}
\multiput(10,100)(60,0){4}{\line(1,-1){20}}
\multiput(39,20)(60,0){3}{\line(1,1){20}}
\multiput(9,40)(60,0){4}{\line(1,-1){20}}
\multiput(39,80)(60,0){3}{\line(1,1){20}}
\multiput(9,100)(60,0){4}{\line(1,-1){20}}
\multiput(41,20)(60,0){3}{\line(1,1){20}}
\multiput(11,40)(60,0){4}{\line(1,-1){20}}
\multiput(41,80)(60,0){3}{\line(1,1){20}}
\multiput(11,100)(60,0){4}{\line(1,-1){20}}
\thinlines
\multiput(40,70)(60,0){3}{\line(1,-1){20}}
\multiput(10,50)(60,0){4}{\line(1,1){20}}
\put(29,13){$\ldots$}
\put(29,73){$\ldots$}
\put(59,43){$w_{2n-1}$}
\put(59,103){$\hat w_{2n-1}$}
\put(89,13){$w_{2n}$}
\put(89,73){$\hat w_{2n}$}
\put(119,43){$w_{2n+1}$}
\put(119,103){$\hat w_{2n+1}$}
\put(149,13){$w_{2n+2}$}
\put(149,73){$\hat w_{2n+2}$}
\put(179,43){$\ldots$}
\put(179,103){$\ldots$}
\put(225,20){$S$}
\put(225,80){$\hat{S}$}
\end{picture}\end{center}
\small\caption{Arrangement of the operators $w_n$ and $\hat w_n$ on
the space-time lattice. }
\label{saw}
\end{figure}

Following \cite{Faddeev:1994} let us give a brief description of the \DQSGM.
The classical model can then be obtained in an appropriate limit.
Let $\q$ be a complex number,
and ${\cal A}_L(\q)$ be an algebra of power series\footnote{
We shall assume these series to be semi-infinite, i.e, the powers of 
$w$ to be bounded from below, but not necessarily non-negative. 
The multiplication of such series
is well defined.} 
in variables  
$w_n$, $n=0,\ldots,2L-1$, $L\ge2$, obeying the following relations
\be\begin{array}{lll}
w_n w_{n-1} &= \q^2 w_{n-1} w_n, \qquad&\forall n,\\
w_n w_m &=  w_m w_n, \qquad &|m-n|\ge 2,
\end{array}\label{algebra}\ee
with the periodicity condition $w_{n+2L}=w_n$.
The variables $w$ constitute the set of the dynamical variables of the
model at any fixed value of time. 
The evolution on one step of the discrete time 
acts as an automorphism $\tau$ of the algebra ${\cal A}_L(\q)$
\be 
\tau: {\cal A}_L(\q) \lra {\cal A}_L(\q),
\ee
such that 
\be\begin{array}{lll}
\hat w_{2n}&=\tau(w_{2n})&=f(\q w_{2n-1}) \,w_{2n}\, f(\q w_{2n+1})^{-1},\\
\hat w_{2n-1}&=\tau(w_{2n-1})&= f(\q \hat w_{2n-2}) \,w_{2n-1}\,
f(\q \hat w_{2n})^{-1}. 
\end{array}
\label{evolution}\ee 
Here and below we use the notation $f(x)^{-1}=1/f(x)$. 
The function $f$ has the form 
\be
f(x)=f(\kappa^2,x),\qquad f(\l,x)={1+\l x\over\l+x}\label{define-f}
\ee
where $\kappa^2$ is a fixed parameter of the model. One can easily
show  that the variables $\hat w_n$ satisfy the same relations (\ref{algebra})
as $w_n$.
To get a geometric picture of the equations (\ref{evolution}) 
consider the square lattice  drawn diagonally as 
 in Fig.~\ref{saw}, with $L\ge2$ sites per row and impose periodic
boundary conditions in the horizontal (spatial) direction. 
The time axis is assumed to be directed upwards. Now assign the 
variables $w_n$, $n=0,\ldots,2L-1$, to the sites belonging
to some  horizontal ``saw'' $\cal S$ as shown in Fig~1. 
In the same way assign 
the  variables $\hat w_n$ 
to the saw $\hat {\cal S}$ shifted 
by one time unit from the initial saw $\cal S$.

\begin{figure}[hbt]\begin{center}
\begin{picture}(100,100)
\put(5,35){\line(1,1){10}}
\put(25,55){\line(1,1){20}}
\put(55,85){\line(1,1){10}}
\put(35,5){\line(1,1){10}}
\put(55,25){\line(1,1){20}}
\put(85,55){\line(1,1){10}}
\put(5,65){\line(1,-1){10}}
\put(25,45){\line(1,-1){20}}
\put(55,15){\line(1,-1){10}}
\put(35,95){\line(1,-1){10}}
\put(55,75){\line(1,-1){20}}
\put(85,45){\line(1,-1){10}}
\put(15,49){$\w_L$}
\put(45,19){$\w_D$}
\put(45,79){$\w_U$}
\put(75,49){$\w_R$}
\end{picture}\end{center}
\caption{Labeling  of the four $w$'s around an elementary face. The
indices stand for ``up'', ``down'', ``left'' and ``right''.}
\label{face}
\end{figure}
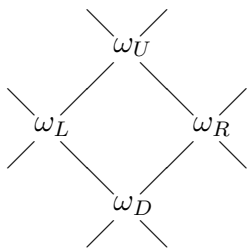

With such arrangement  
both  evolution equations (\ref{evolution}) connect four $w$'s around 
a single square face of the lattice and can be written in a universal
form 
\be\begin{array}{l}
w_U=f(\q w_L) \,w_D\, f(\q w_R)^{-1},\\
w_L w_D=\q^{-2} w_D w_L, \qquad w_D w_R=\q^{-2} w_R w_D,
\end{array}\label{evolution-simple} 
\ee 
where the variables $w$ are labeled as in Fig.~\ref{face}.

The algebra (\ref{algebra}) has two Casimir elements 
\be
C_1 = \prod^{L-1}_{n=0} w_{2n+1} \ , \qquad
C_2 = \prod^{L-1}_{n=0} w_{2n} \ .  \label{casimirs}
\ee
which commute with all $w$'s and are preserved by the evolution 
automorphism $\tau$. Here we restrict ourselves to the case
\be
C_1=C_2=1.\label{constraints}
\ee
This leaves $2L-2$ independent generators, say, $w_0,w_1,\ldots,w_{2L-3}$.

The quasiclassical limit of the model corresponds to the case
$q\rightarrow1$. The algebra (\ref{algebra}) is then replaced by the
Poisson algebra, where 
\be
\{w_n,w_{n-1}\}=2 w_n w_{n-1},\qquad\forall n,
\ee
but all other brackets $\{w_n,w_m\}$ vanish. The equation
of motion (\ref{evolution-simple})  becomes 
\be
w_U=w_D f(w_L)/ f(w_R).\label{classical} 
\ee
One can show that in the appropriate continuous limit the equation 
(\ref{classical}) reduces to \fla{SGE} 
 when $\kappa^2\rightarrow0$. For the details see \cite{Faddeev:1994}.

We will say that the evolution $\tau$ is Hamiltonian if there exists
an invertible operator ${\bf U}$ such that
\be
\tau(a) = {\bf U}^{-1}a{\bf U}.   \label{define-U}
\ee
Let $r(\l,x)$, $\l\in {\Bbb C}$, be a solution of the following 
$\q$-difference equation
\be
r(\l,\q^2 x)=f(\l,\q x) r(\l,x),\label{define-r}
\ee
where the function $f$ is given by (\ref{define-f}). 
Define the operator ${\bf U}$ as follows
\be
{\bf U}
=\prod_{k=0}^{L-1}r(\kappa^2,w_{2k})\prod_{k=0}^{L-1}r(\kappa^2,w_{2k+1}).
\label{formof-U}
\ee
Applying (\ref{define-U}) for  $a=w_n$ and using relation
(\ref{define-r}) one can easily see that (\ref{define-U}) reduces then
to the evolution equations (\ref{evolution}). The calculations are
rather trivial due to the locality of the commutation relations among $w$'s.

To proceed further we need to specify an actual realization of the Hilbert
space of the model, supporting the commutation relations
\fla{algebra}, where the equation \fla{define-r} is well
defined and has a unique solution. At the moment there is no 
classification of representations of this algebra 
suitable for this purpose. Only two cases are known so far. 
The first case is connected with non-compact modular representations 
of \fla{algebra} and the Faddeev-Volkov model
\cite{FV:1993,Faddeev:1994} (see also 
\cite{FKV:2001, Kharchev:2002, Volkov:2005,
Bytsko:2006}). Recently \cite{BMS07a,BMS07b} 
this model was solved exactly and  
shown to be related to the discrete analog 
of the Riemann mapping theorem in two dimensions.
For further details we refer the reader 
to the already mentioned papers \cite{BMS07a,BMS07b} 
and to \cite{BazCP}. 

In the present paper we consider the second known case when the
equation \fla{define-r} is well defined. It is 
when the parameter $\q$ is a root of unity. Originally this case
was  studied in \cite{BBR}.

\subsection{The model at roots of unity.}
Here we are continuing the study of the \DQSGM\ in the limit  when 
$\q^2$ approaches a root of unity. As it was shown in \cite{BBR} the model in
this case can be viewed as a (finite dimensional) integrable quantum system
on top of an integrable classical one.

Let $(-\q_0)$ be a primitive $N$-th root of $-1$, then $\q_0^2$ will always
be a primitive $N$-th root of $1$, 
\be 
(-\q_0)^N=-1,\qquad \q_0^{2N}=1, \qquad N\ge 1.\label{define-roots}
\ee

When $\q=\q_0$ the algebra ${\cal A}_L(\q_0)$ has a (commutative)
center ${\cal Z}({\cal A}_L(\q_0))$ generated by
the elements $w_n^N$.
It is easy to obtain the evolution equations for these 
commuting variables. 
Taking the $N$-th power of both sides of \fla{evolution-simple} for
$\q=\q_0$ one obtains 
\be
w_U^N=w_D^N\,f(\kappa^{2N},w_L^N)/f(\kappa^{2N}, w_R^N),\label{classical-N}
\ee 
by using \fla{define-roots} and the simple identity
\be
f(\kappa^{2N},x^N)=\prod_{j=0}^{N-1}f(\kappa^2,\q_0^{2j+1}),
\ee
where $f(\l,x)$ is defined in \fla{define-f}. Apart from the trivial 
replacement of $\kappa^2$ by $\kappa^{2N}$ the $N$-th powers of $w$ 
obey the same evolution equation as the one 
of the {\it classical}  discrete Sine-Gordon 
model \fla{classical}. In particular, when $N=1$, $\q=1$, formula 
\fla{classical-N} exactly reduces to \fla{classical}  as it, of course,
should. 

The quantum evolution of the model takes place in a finite dimensional
factor  of the algebra ${\cal A}_L(\q_0)$ over its center 
${\cal Z}({\cal A}_L(\q_0))$. In fact,  let
$\a=\{\a_0,\ldots,\a_{2L-3}\}$ be a set of nonzero complex numbers and 
$I_{\a}$ be an ideal generated by $(w_n^N-\a_n)$, $n=0,\ldots,2L-3$.
Then the factor ${\cal A}_L(\q_0)/I_{\a}$ for any $\a$ 
is isomorphic to a finite 
dimensional algebra generated by elements $X_i,Z_i$, $i=1,\ldots,L-1$
obeying the  following relations
%\begin{numlet}
%\renewcommand{\arraystretch}{2}
%\setlength{\jot}{2mm}
\be\begin{array}{ll}
X_i Z_i =\q^2_0\, Z_i X_i, \qquad &X_i^N=Z_i^N=1,\label{PQXZ-a}\\
X_i Z_j =Z_j X_i, \qquad &i\ne j, \label{PQXZ-b}
\end{array}\label{XZ-algebra}
\ee
%\renewcommand{\arraystretch}{1.5}
%\end{numlet}
The explicit isomorphism is achieved by the formulae
\be 
\begin{array}{ll}
w_{2n}=\e^{P_{n+1}}X_{n+1}, \qquad &n=0,\ldots,L-2,\\
w_{2n+1}=\e^{Q_{n+1}-Q_{n+2}}Z^{-1}_{n+1}Z_{n+2}, \qquad &n=0,\ldots,L-3,\\
w_{2L-3}=\e^{P_{2L-3}}Z_{L-1}^{-1}\ ,
\end{array}\label{generators}
\ee
where for later convenience we have parametrized the $\a$'s through
a set of new variables $P_i,Q_i$, $i=1,\ldots,L-1$ as
\be 
\begin{array}{ll}
\a_{2n}=\e^{P_{n+1}}, \qquad &n=0,\ldots,L-2,\\
\a_{2n+1}=\e^{Q_{n+1}-Q_{n+2}}, \qquad &n=0,\ldots,L-3,\\
\a_{2L-3}=\e^{P_{2L-3}}.
\end{array}\label{alphas}
\ee
The remaining two dependent generators $w_{2L-2}$ and $w_{2L-1}$ 
can be expressed from
\fla{generators}, \fla{casimirs} and \fla{constraints}.

One  can introduce a Poisson algebra structure
on ${\cal Z}({\cal A}_L(\q_0))$ with its Poisson  
action by derivations on the whole ${\cal A}_L(\q_0)$. In fact, this
structure naturally emerges from \fla{algebra} in the limit
$\q\rightarrow \q_0$. 
First, recall that the Poisson algebra $P$ is by definition a
commutative algebra with the structure of a Lie algebra, defined 
by a bracket  $\{\ ,\ \}$ and the rule 
\be
\{a,bc\}=\{a,b\}c+b\{a,c\},\label{rule}
\ee
for any $a,b,c\in P$,  which determines a  distributive 
action of the bracket on the (commutative) product. Next, we can
identify the algebras ${\cal A}_L(\q)$ for all $q$ as vector spaces 
of formal power series in generators by choosing some definite way of
their ordering. For convenience set
\be \q=\e^{h}\q_0 \label{q-h}
\ee
where $h$ is a formal variable.
Denoting by $a\cdot b$ the product of two elements in 
${\cal A}_L(\q)$ we can define the bracket 
\be 
\{a,b\}=\lim_{h\rightarrow0}{a\cdot b-b\cdot a\over h}\label{bracket}
\ee
when at least one of the elements $a$ and $b$ belongs to ${\cal
Z}({\cal A}_L(\q_0))$. Using \fla{algebra}
it is easy to compute the Poisson brackets \fla{bracket}
 between the generators $w_n^N$
of ${\cal Z}({\cal A}_L(\q_0))$  and their action on $w_n$:
\be\begin{array}{ll}
\{w_m^N,w_n^N\}=&4N^2(\d_{m+1,n}-\d_{m,n+1}) w_m^N w_n^N\\
\{w_m^N,w_n\}=&2N(\d_{m+1,n}-\d_{m,n+1}) w_m^N w_n
\end{array}\label{action}
\ee
where 
\be
\delta_{m,n}=\left\{\begin{array}{ll}1,&m=n \pmod {2L},\\
                                        0,&\mbox{otherwise.}
                     \end{array}\right.
\ee
The Poisson action of ${\cal Z}({\cal A}_L(\q_0))$ on the whole 
${\cal A}_L(\q_0)$ is extended from \fla{action} by the same rule
\fla{rule}
where now  $a\in{\cal Z}({\cal A}_L(\q_0))$ and $b,c\in{\cal A}_L(\q_0)$
(of course, the multiplication there should now be assumed non-commutative).

For later use introduce the following notation 
\begin{equation}
\exp(a)\circ b=\sum_{n\ge 0}
{1\over n!}\underbrace{\{a\cdots\{a}_n,b\}\cdots\}
\label{exp-action}
\end{equation}
for the Poisson action of $\exp(a)$
on $b$, where $a\in{\cal Z}({\cal A}_L(\q_0))$ and $b\in{\cal A}_L(\q_0)$.
It follows from \fla{bracket} that
\begin{equation}
\exp\left({a\over h}\right)\cdot b\cdot
\exp\left(-{a\over h}\right)=
\exp(a)\circ b+O(h)
\label{adjoint-action}
\end{equation}
If we use the parametrization \fla{generators}  then 
\fla{action} reduces to the only non-trivial brackets
\be
\{Q_i,P_j\} = \d_{ij}
\ee

The standard quantization of the  
Poisson structure described above 
gives a realization of the algebra ${\cal A}_L(\q)$, 
$\q=\e^h\q_0$,  
within a wider algebra of formal power series
\be {\cal X}(P,Q,X,Z,h)=\hbox{\rm (series in $P_i$,$Q_i$,$X_i$,$Z_i$,
$h$)}
\ee
in non-negative powers of $P_i$, $Q_i$, $X_i$, $Z_i$, 
$i=0,\ldots,L-1$ and $h$, where $X_i$, $Z_i$ are defined in \fla{XZ-algebra}
and $P_i$,  $Q_i$ satisfy the relations
\be\begin{array}{l}
{[ Q_i ,P_j ]}=2h\d_{ij}, \qquad [P_i,h]=[Q_i,h]=0,\\
{}{[P_i,X_j]}=[P_i,Z_j]=[Q_i,X_j]=[Q_i,Z_j]=0.
\end{array}\label{PQ-algebra}
\ee
The generators $w_n$ are given 
by \fla{generators} where the elements 
$P_i$,$Q_i$ now satisfy \fla{PQ-algebra}.
This realization of ${\cal A}_L(\q)$
is most suitable for the  study  
of the limit $\q\rightarrow \q_0$, $\q_0^{2N}=1$.

Note also that  when $L=2$ the algebra
${\cal A}_L(\q)$ has only two independent 
generators $w_0$ and  $w_1$ which we denote as $v$ and $u$ respectively.
From \fla{algebra} we have
\be
uv=\q^2vu
\ee
while the realization \fla{generators} contains only one pair of the 
$P,Q$ and one pair of the $X,Z$ operators 
\be
u=\e^Q Z^{-1}\qquad v=e^P X\label{uv}
\ee
where 
\be
XZ =\q^2_0\, Z X, \qquad X^N=Z^N=1,\label{XZ-alg}
\ee
\be
[Q,P]=2 h\qquad [P,h]=[Q,h]=0,\label{PQ-alg}
\ee
where $h$ and $\q$ are related by \fla{q-h}.

\subsection{The integrability of the model.}
\label{integrability}

In this section we briefly address  the matters related to the integrability 
of the \DQSGM\ which so far have not been discussed. The details of
the calculations can be found in \cite{BBR}.

Let ${\mathsf W}_\q$ be the Weyl 
algebra generated by the invertible elements $U$ and $V$,
$$
{\mathsf W}_\q:\qquad\qquad UV = \q\, VU\ ,    \label{UV-pair}
$$
where $\q\in{\mathbb C}$. Further, let $R(\l)$ be the $R$-matrix of the
6-vertex model 
 acting in $\Bbb C^{2}\otimes \Bbb C^{2}$ 
\be\begin{array}{ll}
R(\l) & = (\l \q-\l^{-1}\q^{-1})(e_{11}\otimes e_{11}+e_{22}\otimes e_{22}) \\
     &\quad + (\l-\l^{-1})(e_{11}\otimes e_{22}+e_{22}\otimes e_{11}) \\
     &\quad + (\q-\q^{-1})(e_{12}\otimes e_{21}+e_{21}\otimes e_{12}) \ ,
\end{array}   \label{six-vertex}
\ee
where $e_{ij}$ is the matrix unit and $L(\l)$
be an $L$-operator acting in $\Bbb C^2\otimes {\mathsf W}_\q$ 
\be
L(\l) = \pmatrix{U & -\l V\cr
          \l V^{-1}  &      U^{-1}\cr }.\label{L-op}
\ee
These operators satisfy two \YBE s 
\be
R_{12}(\l)R_{13}(\l\mu)R_{23}(\mu) =
R_{23}(\mu)R_{13}(\l\mu)R_{12}(\l)     \label{ybe-a}
\ee
and
\be
R_{12}(\l)L_1(\l\mu)L_2(\mu) = L_2(\mu)L_1(\l\mu)R_{12}(\l)   \label{ybe-b}
\ee
which are written in the standard notations.
As usual, define the 
transfer matrix 
\be
t(\l) = {\rm trace}(L_{0}(\l\kappa)L_{1}(\l/\kappa) L_{2}(\l\kappa)\cdots
       L_{2L-1}(\l/\kappa)) ,    \label{define-t}
\ee
for a periodic chain of the length $2L$ 
with alternating rapidity variables $\l\kappa$, $\l/\kappa$,
$\l\kappa$, etc., where $\kappa$ is a constant. 
This transfer matrix acts in a direct product 
${\mathsf W}_\q^{\otimes 2L}$ of $2L$ copies of the algebra ${\mathsf W}_\q$.

  As a consequence of \fla{ybe-b} the
transfer matrices \fla{define-f} with different $\l$ form a commuting family.
From the definitions
 \fla{L-op} and \fla{define-t} it is trivial to see that $t(\l)$ is a
polynomial of the degree $2L$ in $\l^2$. What is less trivial, the 
operators $U_i$, $V_i$, $i=0,\ldots,2L-1$, associated with
different sites of the chain, enter coefficients of this polynomial 
only trough $2L$ elements of the form (remind that we work with 
periodic boundary conditions)
\be
w_n=U_nV_n^{-1}U_{n+1}V_{n+1},\label{combinations}
\ee
which obey the commutation relations of the algebra \fla{algebra}!
The proof of this statement can be found in \cite{BBR}.
The next remarkable fact is that the transfer matrix \fla{define-t} 
commutes with the evolution operator \fla{define-U} and, therefore, it
generates commuting integrals of motion of the
\DQSGM .

To see this let us search for  another $R$-matrix acting in
${\mathsf W}_\q\otimes {\mathsf W}_\q$ which intertwines two $L$
operators in the ``quantum'' 
space. In fact,  this $R$-matrix has already appeared in Sect.~1.  
Assume that there exists a well defined element 
$r(\l,u)\in {\mathsf W}_\q\otimes {\mathsf W}_\q$, where 
\be
u=UV^{-1}\otimes UV,\label{local-w}
\ee
and $r(\l,x)$ is the solution of the $\q$-difference equation \fla{define-r}.
\be
r(\l,\q^2 x)={1+\q \l x\over \l+ \q x} r(\l,x), \label{q-diff}
\ee
Then, it is easy to show that
\be
r(\l,u)(L(\l\mu)\otimes L(\mu))=
(L(\mu)\otimes L(\l\mu))r(\l,u)\label{ybe-c}
\ee
With an account of \fla{define-U}, \fla{combinations} and
\fla{local-w} this fact immediately implies the required commutativity 
of the evolution operator of \fla{define-U} and the transfer matrix 
\fla{define-t} 
\be 
[t(\l),{\bf U}]=0
\ee

At this point we would like to note that so far we used only the defining 
relation \fla{q-diff} for the $R$-matrix $r(\l,x)$ rather than its
explicit form.  This concerns the \YBE\ \fla{ybe-c} and the
equivalence of the two forms \fla{evolution} and \fla{define-U} of the  
equations of motion in the model. An explicit form of $r(\l,x)$ 
for the case when $\q^2$ is approaching a root of unity will be given in
the next Section.

\nsection {The \YBE\ for $r(\l,x)$.}
\subsection{General analysis.}
The reader may have noticed that the sequence of \YBE s
\fla{ybe-a}, \fla{ybe-b}, \fla{ybe-c} in the
previous Section  which starts 
from the six-vertex model is very much resembling the one that
appeared  in the analysis   of the chiral Potts model in ref. \cite{BS90}. 
In fact, our  considerations here are quite similar but 
more general than in \cite{BS90} and  the 
relations with the Chiral Potts model will soon be very explicit.

The final point in the descending sequence of the \YBE s mentioned
above is the equation for $R$-matrix $r(\l,x)$ itself.
The form of \fla{ybe-c} requires  that it should be written 
as  the ``braid relation''
\be 
r(\lambda,u)r(\lambda\mu,v)r(\mu,u)=r(\mu,v)r(\lambda\mu,u)r(\lambda,v),
\label{YBE}\ee
where  $v$ and $u$ denote respectively any two successive generators $w_n$ and
$w_{n+1}$ from \fla{local-w} which form a Weyl pair
\be
 uv=\q^2 vu. \label{uv-rel}
\ee
Here $\l,\mu\in\Bbb{C}$ are the (multiplicative) rapidity variables.
It might be tempting to assume that the two relations 
\fla{q-diff} and \fla{uv-rel} are sufficient to imply the 
\YBE\ \fla{YBE}.
As we shall see below this statement is not in general  true.
However, it would  be quite interesting to understand how far one could 
advance in proving \fla{YBE} using  those two relation only.

Combining (\ref{q-diff}) and \fla{uv-rel} one obtains
\be
%\begin{array} {ll}
r(\l,u)\,v = v\, {1+\q\l u\over \l+\q u} \, r(\l,u),\qquad
r(\l,v)\,u = u\, {\q\l+v\over \q+\l v} \, r(\l,v).
%\end{array}
\label{ruv-equation}
\ee
Denote the left and right hand sides of the \YBE\ (\ref{YBE}) 
as $\Phi_L$ and $\Phi_R$  respectively. Then using the relations
(\ref{uv-rel}),(\ref{ruv-equation}) only one can show that for both
$\Phi=\Phi_L$ or $\Phi=\Phi_R$ 
\be
\Phi\,u=u g(u,v,\q) \Phi, \qquad \Phi\,v= v\, \Phi (g(v,u,\q^{-1}))^{-1},
\label{comm}
\ee
where
\be
g(u,v,\q)=(\q\l^2\mu+ \q^2\l\mu u +v+q\l v u)
(\q\l+ \q^2 u+\l\mu v+\q^2\l^2\mu v u)^{-1}.
\label{define-g}\ee
The calculations are elementary
but a bit tedious. We present them in Appendix~A.
Fortunately, we have to derive only one of the equations in \fla{comm} 
since they follow from one another under the action of the 
automorphism 
\be
r(\l,x) \lra (r(\l,x))^{-1},\qquad
u \llra v, \qquad \q \lra \q^{-1},\label{autho}
\ee
which leaves (\ref{uv-rel}),(\ref{ruv-equation})
unchanged. There are also additional ``exchange'' relations which follow from
\fla{comm} and the fact that $r(\l^{-1},x)^{-1}$ satisfies the same
$\q$-difference equation \fla{q-diff} as $r(\l,x)$
\be
\Phi\, u =u\, \Phi ({\overline g}(u,v,\q))^{-1},
\qquad \Phi\,v= v\, {\overline g}(v,u,\q^{-1})\Phi.
\label {comm-a}\ee
As above, these relations are valid 
for both $\Phi=\Phi_L$ or $\Phi=\Phi_R$. The function
${\overline g}(u,v,\q)$ is given by the RHS of \fla{define-g} with 
$\l$ and $\mu$  replaced by $\mu^{-1}$ and $\l^{-1}$ respectively.
The relations (\ref{comm}) and \fla{comm-a} immediately imply that
\be 
[\Psi, u] = [\Psi, v] = 0. \label{result}
\ee
where 
\be 
\Psi=\Phi_L\Phi_R^{-1}\label{ratio}
\ee
Apparently, this is the most general consequence for   the \YBE\
\fla{YBE} which one 
can obtain from (\ref{uv-rel}),(\ref{ruv-equation}) without further
requirements on the properties of the $r(\l,x)$.  
Therefore, if the representation of \fla{algebra} were such that
the commutativity with the elements $u$ and $v$ in \fla{result} implies that 
the quantity $\Psi$ is proportional to the unit operator, then \YBE
\fla{YBE} would be satisfied and, at the same time, equation
\fla{define-r} would have a unique solution.  

Let $\q=\e^h \q_0$, where $h$ is a formal
variable and $\q_0$ is defined in \fla{define-roots}. 
The most appropriate way to study \fla{define-r} 
when $\q^2$ approaches a root of unity, 
$h\rightarrow0$, is to 
consider
$r(\l,x)$ and, hence,  the left and right hand sides of 
equation \fla{YBE} as asymptotic expansions
in $h$.
Define the 
Euler dilogarithm function as
\be
Li_2(x)=-\int_0^x{\log (1-t)\over t}dt.
\label{Euler}
\end{equation}
The \YBE\  \fla{YBE} involves only one Weyl pair
\fla{uv-rel} which we parametrize as in \fla{uv}. Then it is 
convenient to rewrite \fla{q-diff} as 
\be
r(\l,\e^{T+h}\q_0z)={1+ \l x\over \l+  x}r(\l,\e^{T-h}z/\q_0).\label{q-diff-a}
\ee
Here we set $x=\e^Tz$, where $z$ is a formal variable such that 
$z^N=1$.
Next, impose the following normalization condition on $r(\l,x)$
\be
r(1,x)=1.\label{norm}
\ee
Then equation \fla{q-diff} uniquely determine the 
asymptotics expansion of $r(\l,x)$ in the form
\be 
r(\l,\e^Tz) = \exp\left\{c_{-1}(T,z)h^{-1}+c_0(T,z) + c_1(T,z)h 
+\ldots\right\}, 
\label{h-exp}
\ee
where the coefficients $c_k(T,z)$ are series in $T$ and $z$. In principle,
all these
coefficients can be calculated explicitly from \fla{q-diff-a}. In particular,
retaining only the first two terms in \fla{h-exp} one obtains 
\be
r(\l,x)=\exp\left({H(\l^N,x^N)\over
N^2 h}\right)\overline r(\l,x)(1+O(h))
\label{asymp}
\end{equation}
where $x=\e^T z$,
\begin{equation}
H(a,b)=-{1\over 2}\left\{Li_2(-ab)
+Li_2(-a/b)
+\log^2 b+\pi^2/6\right\}
\label{define-H}
\end{equation}
and
\begin{equation}
\overline r(\l,x)
=\left({\l^N+x^N\over 1+\l^Nx^N}\right)^{(N-1)/2N}
\quad\prod^{N-1}_{j=0}
\left({1+\l xq_0^{2j+1}\over
\l+xq_0^{2j+1}}\right)^{j/N}~.\label{rbar}
\end{equation}
The branches of the function $Li_2(x)$ are assumed to be 
chosen such that $H(1,x)=0$,
so that the normalization condition \fla{norm} is satisfied.
The expansion \fla{asymp} can be easily deduced from formula
(3.11) of ref. \cite{Bazhanov:1995jpa}.

\subsection{The case $\q\rightarrow1$.}

First, consider the case $\q\rightarrow 1$. This corresponds 
to $N=1$, $\q_0=1$, $\q=\e^h$ in our previous notations.
Eq.\fla{uv} now reads
\be
u=\e^Q,\qquad v=\e^P. \label{uvPQ}
\ee
It follows from  \fla{h-exp} that both elements $r(\l,u)$ and $r(\l,v)$ 
have the form of exponentials 
\be \exp({\cal X}/h),\label{expo}
\ee
where ${\cal X}$
is a series 
\be
{\cal X}=\sum_{k,\ell,m=0}^{\infty}c_{k\ell m}h^k Q^\ell P^m, \label{pq-series}
%{\cal X}=\hbox{\rm (series in $P$,$Q$,$h$)}
\ee
in non-negative powers of $P$, $Q$ and $h$.
Of course, for $r(\l,u)$ and $r(\l,v)$ the above series do not contain
mixed terms (they involve only pure powers of $P$ or $Q$). 
However, below we will consider more general series (\ref{pq-series}).
We will always assume them to be ``normally ordered'' such 
that all $P$'s  appear to the right from $Q$'s.
The product of such two exponentials is again an exponential of the same 
form. Indeed, from the Campbell-Hausdorff formula we have
\be
\e^{{\cal X}/h}\e^{{\cal Y}/h}= \e^{{\cal Z}/h},
\label{xyz}\ee
\be
{\cal Z}={\cal X}+{\cal Y}+{1 \over 2 h}[{\cal X},{\cal Y}]+
{1 \over 12 h^2}\left\{[{\cal X},[{\cal X},{\cal Y}]]+
[{\cal Y},[{\cal Y},{\cal X}]]\right\}+\ldots
\label{CH-series}\ee
where for each term of the last series the number of commutators 
coincides with the   
power of the parameter 
$h$ in the denominator. If both ${\cal X}$ and ${\cal Y}$
are of the form \fla{pq-series}, then due to \fla{PQ-alg} each commutator
produces one extra power of $h$ in the numerator so that all  negative
powers of $h$ in \fla{CH-series} exactly cancel out. Therefore ${\cal Z}$ 
is a series of the form \fla{pq-series} as well. It should be noted that each 
coefficient $c_{k\ell m}$ therein will now be an infinite 
 numerical series in the  coefficients of ${\cal X}$ and 
${\cal Y}$. For the moment let us assume that all these numerical series
converge and, hence, ${\cal Z}$ is well defined. 
Repeatedly applying 
\fla{xyz} one  concludes that the 
ratio $\Psi$ of the left and right hand sides of the \YBE\ 
in \fla{ratio} is also an exponential of the form \fla{expo}. 
Then it is easy to show that the most general form of $\Psi$ 
satisfying \fla{result}, \fla{uvPQ}  is 
\be
\Psi=\hbox{\rm const }\exp\left\{\ib \pi (m P+n Q)/h\right\} ,
\ee
where $m$ and $ n$ are arbitrary integer constants. 
Finally setting the spectral 
parameters $\l=\mu=1$ in \fla{YBE} and using \fla{norm} 
one concludes that $\Psi\equiv1$. Since we worked with exponentials of 
asymptotic series in $h$ 
this implies that the \YBE\ \fla{YBE} is satisfied as an equality of such 
exponentials   in any given order 
in $h$. 

It should be noted that the series in $h$ in 
the exponents of \fla{h-exp} and 
\fla{xyz}
can be consistently truncated at any level. In particular,
setting $N=1$ and taking into account  
only the two leading terms in \fla{h-exp} one obtains the following
\YBE\ 
\be 
r(\lambda,u)r(\lambda\mu,v)r(\mu,u)=r(\mu,v)r(\lambda\mu,u)r(\lambda,v)
(1+O(h)),\label{truncated} 
\ee
where
\be
r(\l,x)=\e^{H(\l,x)/h}(1+O(h))\label{asymp-b}.
\ee
Writing \fla{truncated} in full and substituting \fla{uvPQ} one obtains
\be\begin{array}{l} 
\exp(H(\l,\e^Q)/h)\,\exp(H(\l\mu,\e^P)/h)\,\exp(H(\mu,\e^Q)/h)\\
=\exp(H(\mu,\e^P)/h)\,\exp(H(\l\mu,\e^Q)/h)\,\exp(H(\l,\e^P)/h)
(1+O(h))
\end{array}\label{cYBE}
\ee
Introducing a special notation for the leading term of the Campbell-Hausdorff
composition in \fla{xyz} 
\be
\exp\left({{\cal X}\over h}\right)
\exp\left({{\cal Y}\over h}\right)= O(1)\exp\left({{\cal X}\ast{\cal Y}\over 
h}\right),
\qquad h\rightarrow 0,
\ee
we can rewrite  
the equality of the singular-in-$h$ exponents in \fla{truncated}
as
\be
H(\l,u)\ast H(\l\mu,v)\ast H(\mu,v)=
H(\mu,v)\ast H(\l\mu,u)\ast H(\l,u).\label{singular}
\ee
Below we show that this formula is equivalent to some identity
for the dilogarithm function \fla{Euler}. 

To complete the 
proof of the \YBE\ we have to ensure that the abovementioned
numerical series involved in the product of the $R$-matrices converge.
Apparently this can be done ``order by order'' in perturbations in $h$ 
using a special 
structure of the Campbell-Hausdorf series \fla{CH-series} and the fact
that the numerical coefficients there rapidly decrease (see, e.g.,
\cite{Bourbaki}, chapter 2, \S 6.4).
We will not analyze this problem in its general setting here,
but restrict our considerations to the truncated equation \fla{truncated}. 
In this case we are able not only to show that products
of the $R$-matrices in \fla{truncated} exist  but also explicitly calculate
normal symbols of the two sides of \fla{truncated} 
up to the order of $O(h^0)$. The details  of the 
calculations are presented below.

\subsection{The 12-term dilogarithm identity}

Let $A$ be an operator given 
as a power series in $\hat P$ and $\hat Q$ satisfying\footnote{
In this subsection we will denote the operators $P$ and $Q$ from
\fla{PQ-alg} as $\hat P$ and $\hat Q$ and use the letters $P$ and $Q$
for commuting arguments of the normal symbols. We hope this will not
cause any confusion.}
\fla{PQ-alg} 
\be
A=\sum_{k,l=0}^{\infty}a_{kl} {\hat Q}^k {\hat P}^l \label{operator-A}
\ee
ordered such that all $\hat P$'s are placed on the right from $\hat Q$'s. 
We will 
call the series ordered in this way as normal form of the operator $A$.
As usual the 
 normal symbol $:A(Q,P):$ of the operator $A$ is defined by the same 
series \fla{operator-A} as its normal form but with  $\hat P$ and
$\hat Q$ replaced by commuting variables $P$ and $Q$ respectively.

To calculate the normal symbol of the product of two different operators
one has to use a simple identity  for the normal symbol of an
elementary (unordered) monomial
\be
A_{kl}=\hat P^k\hat Q^l
\ee
\be
:A_{kl}(Q,P):=(-2h)^{k+l}\exp(PQ/{2h}){\partial^k \over \partial Q^k}
{\partial^l  \over \partial P^l}\exp(-PQ/{2h})
\label{ident}
\ee
which can be easily proved from \fla{PQ-alg} by induction. Following
\cite{Berezin} we can write the RHS of  \fla{ident} in an integral form 
\be
(\ref{ident})= 
\int   {d P' d Q'\over 2\pi h}\,\exp((Q-Q')(P-P')/2h){P'}^k {Q'}^l
\ee
assuming $h$ to be negative real and considering $P'$ and $Q'$ as 
complex conjugated variables $P'=(Q')^*\in{\Bbb C}$. The integral is
assumed to be over the whole complex plane. Then the normal symbol of the 
product of two operators $A$ and $B$ can be written in a compact way
\be
:AB(Q,P):=\int
{dQ' dP'\over2\pi h}  \exp((Q-Q')(P-P')/2h) :A(Q,P'):\,:B(Q',P):
\ee
Applying  this for the left hand side of \fla{truncated} one obtains
\be
:\Phi_L(Q,P):=r(\l,\e^Q)\int
 \e^{(Q-Q')(P-P')/2h}\, 
r(\l\mu,\e^{P'})\,r(\mu,\e^{Q'})\,{dQ' dP'\over2\pi h}\label{LHS}
\ee
This integral can be explicitly calculated to within terms of the order
of $O(h^0)$ using the saddle point approximation. A similar treatment of
the quantum five-term identity can be found in \cite{FK}.
Substituting 
\fla{asymp-b} into \fla{LHS} one obtains
\be\begin {array}{l}
:\Phi_L(Q,P):=\displaystyle (\det{\bf M}(x',y'))^{-1/2}\exp(E_L/2h)
(1+O(h))\ ,\\
E_L=
H(\l,x)+H(\l\mu,y')+H(\mu,x')+2\log(x/x')\log(y/y')
\end{array}\label{normsym}
\ee
where 
\be
x=e^Q,\qquad y=\e^P,\qquad x'=e^{Q'},\qquad y'=\e^{P'}, 
\ee 
and the coordinates $P'$, $Q'$ of the saddle point are determined by the 
equations
\be
x={x'(1+\l\mu y')\over \l\mu+y'},\qquad  
y={y'(1+\mu x')\over \mu+x'},\label{variab-a}
\ee
The two by two matrix $\bf M$ coincides (up two a factor of $2h$) with the 
matrix of the quadratic form in the exponent of the 
integrand near the saddle point
\be
{\bf M}(x',y')=\pmatrix{\displaystyle 
{\frac {{\it x'}\,\left (\mu^{2}-1\right )}{\left ({\it x'}+\mu
\right )\left (1+\mu\,{\it x'}\right )}}& 1\cr
1&\displaystyle{\frac {{\it y'}\,\left (\lambda^{2}\mu^{2}-1\right )}
{\left (1+
\lambda\,\mu\,{\it y'}\right )\left ({\it y'}+\lambda\,\mu\right )}}
\cr}
\ee
The equations \fla{variab-a} 
determine two distinct saddle points. Instead of solving these equations 
with respect to $x'$ and $y'$, it is more convenient to regard $x'$ and
$y'$ as
independent variable and consider \fla{variab-a} as a definition of
$x$ and $y$.  In the same way one can calculate the normal symbol of
the RHS of  \fla{truncated}. 
Then for any $x'$ and $y'$ the 
 corresponding saddle point for $:\Phi_R(Q,P):$
is given by
\be
y''={x'(\l\mu x'y' +x' +\l y' +\mu)\over \mu x'y' +\l x' +y'+\l\mu},
\qquad
x''={y'(\l\mu x'y' +x' +\l y' +\l^2\mu)\over \l^2\mu x'y' +\l x' +y'+\l\mu}.
\label{variab-b}\ee
where 
\be
x''=\e^{P''}, \qquad y''=\e^{Q''},
\ee
and $P''$ and $Q''$ are integration variables in the corresponding 
integral for $:\Phi_R(Q,P):$. The expression for $:\Phi_R(Q,P):$ is
obtained from \fla{normsym} 
merely by replacing $x'$ and $y'$ therein with $x''$ and
$y''$ respectively.
Equating the singular-in-$h$ exponents
of the normal symbols we thus obtain the following identity.

\vspace{0.5cm}
{\it The twelve-term dilogarithm identity}.
Let $\l$, $\mu$, $x'$ and $y'$ be 
arbitrary complex numbers and $x$, $y$, $x''$, $y''$ are given by 
\fla{variab-a}, \fla{variab-b}.
Then
\be\begin{array}{l} 
\ \ H(\l,x)+H(\l\mu,y')+H(\mu,x')+2\log(x/x')\log(y/y')\\
=H(\mu,x'')+H(\l\mu,y'')+H(\l,y)+2\log(x/y'')\log(y/x''),\label{twelve}
\end{array}
\ee
where the function $H$ is defined by \fla{define-H}.
The branches of the dilogarithms and logarithms are to be chosen as analytic 
continuation from the case when $\l$, $\mu$, $x'$ and $y'$ are all real and 
positive. 

\vspace{0.5cm}
Since we have shown (by explicit calculations) that products of the
$R$-matrices in the \YBE\ \fla{truncated} do  exist and are well defined 
(to within the required order in $h$) the identity \fla{twelve}
is a corollary of 
our proof of the \YBE\ in the previous subsection.  However, 
is very easy to check directly the dilogarithm identity of this
type 
once it is written. First, computing  partial 
derivatives of the first order in the variables $\l$, $\mu$, $x'$ or 
$y'$  one can 
check that the difference between the LHS and RHS of \fla{twelve}
considered as a function of these four independent variables is a constant.
To check that 
this constant is actually equal to zero one has to set $x'=y'=1$.
The four variable dilogarithm identity \fla{twelve} 
seems to be new, but, of course, can be obtained by a repeated
application of the classical five term identity for the 
dilogarithm function \fla{Euler} (for 
the most comprehensive survey on this subject see \cite{K}).

Note that \fla{variab-a}, \fla{variab-b} imply
\be
x={y''(1+\mu x'')\over \mu+x''},\label{variab-c}\qquad
y={x''(1+\l\mu y'')\over \l\mu+y''}.
\ee
Regarding $\l$ and $\mu$ as constants and $x''$, $y''$ as independent
variables we can write   the RHS of \fla{twelve}
as some function  ${\cal F}_0(x'',y'')$. 
Then the identity \fla{twelve} can be interpreted as 
an invariance of this function under the (bi)rational substitution 
\fla{variab-b} of two variables 
\be
{\cal F}_0(x'',y'')={\cal F}_0(x',y')
\ee
Note that the substitution \fla{variab-b} is an involution, i.e, it reduces
to identity substitution on the second iteration.

Similarly, the order of $O(h^0)$ in \fla{truncated} (which is equivalent to
to the equality of the determinants in \fla{normsym} and in the corresponding
expression for $:\Phi_R(Q,P):$)
gives a 
polynomial invariant 
\be
{\cal F}_1(x'',y'')=(1+\mu x'')(1+\l\mu y'')
\ee
of the substitution \fla{variab-b}
\be
{\cal F}_1(x'',y'')={\cal F}_1(x',y').
\ee
Of course, the last formula  can be readily checked by direct calculations. 

\subsection{The case of a generic root of unity}
Now let $\q_0$ be a primitive $N$-th 
root of $-1$ as defined in \fla{define-roots} and
\be \q=\e^{h}\q_0 
\ee
The generators $u$ and $v$ are now given  by the general formulae
(\ref{uv}-\ref{PQ-alg}) and the leading asymptotics of the $R$-matrix
by \fla{asymp}. 
Our previous  proof of the truncated \YBE\ \fla{truncated} can be
easily modified for this case. One has to generalize the 
series in \fla{pq-series} to series in $P$, $Q$, $X$, $Z$ and $h$
(of course the powers of $X$ and $Z$ will  not exceed $N-1$).
The subsequent arguments  are very similar to those   
of the case $\q\rightarrow1$  and we will not present them here.

Let us rather consider consequences of the \YBE\ \fla{truncated} in
this case. The $R$-matrix \fla{asymp} factorizes into a ``classical''
and a ``quantum'' part.  
The former is the 
singular-in-$h$ exponential belonging to the center of the algebra 
${\cal Z}({\cal A}_{2}(\q_0))$ while the latter belongs to the finite 
dimensional factor of this algebra over its center, which is generated only
by the elements $X$ and $Z$ obeying \fla{XZ-alg}.
It turns out that that the \YBE\ \fla{truncated} remarkably splits  
into two 
separate (quantum) 
\YBE s for the classical and the quantum parts of the $R$-matrix.
The trick is similar to that used in \cite{Bazhanov:1995jpa} 
for the asymptotics
of the quantum five-term identity of ref.\cite{FK}.
The equation for the classical parts is equivalent to
\fla{cYBE} while the one for the quantum parts 
appears to be equivalent (as we shall see in the following sections)
to the star-triangle relation of the $N$-state chiral Potts model.

Substitute \fla{asymp} into \fla{truncated} and move all 
singular-in-$h$ exponentials to the right. 
To do this one has to use formula
\fla{adjoint-action}, since these exponentials in general 
 do not commute with the
elements $\overline r$.
For instance, for the LHS of \fla{truncated} one obtains
\be
\renewcommand{\arraystretch}{2.5}
\begin{array}{l}
\ds\e^{\HT(\l,\e^Q)/h}\,\, {\overline r}(\l,\e^Q Z^{-1})\,\,
\e^{\HT(\l\mu,\e^P)/h} \,\,{\overline r}(\l\mu,\e^{P}X)\,\,
\e^{\HT(\mu,\e^Q)/h}{\overline r}\,\,(\mu,\e^{Q}Z^{-1})=\\
\ds{\overline r}(\l,\e^Q Z^{-1})\biggl(\e^{\HT(\l,\e^Q)}\circ
{\overline r}(\l\mu,\e^{P}X)\biggr)
\biggl(\e^{\HT(\l\mu,\e^P)}\circ
\e^{\HT(\mu,\e^Q)}\circ {\overline r}(\mu,\e^{Q}Z^{-1})\biggr)\times\\
\times \e^{\HT(\l,\e^Q)/h}\e^{\HT(\l\mu,\e^P)/h} 
\e^{\HT(\mu,\e^Q)/h}(1+O(h)),
\end{array}
\renewcommand{\arraystretch}{1.5}
\ee
where
\be{\widetilde H}(a,b)=
N^{-2}H(a^N,b^N)
\label{Htilde}\ee
with the function 
$H$ given by \fla{define-H}.
Perform similar transformations for the RHS of \fla{truncated}.
After that all singular  exponentials in the LHS cancel with those 
in the RHS side due to  \fla{cYBE} (more precisely, one
has to use \fla{cYBE} with the function $H$ replaced by a function  
$\widetilde H$, defined in \fla{Htilde}).  
The remaining equation is that for the quantities of the order of
$O(h^0)$ so that 
one can now set $h=0$. The operators $P$ and $Q$ then become commuting 
variables. More precisely, they become elements of the Poisson
algebra with  the  bracket
\be
\{Q,P\}=2h\label{PQbracket}
\ee
as it was explained in Section~2.2.
As a result we obtain the following ``twisted'' \YBE\ for  the elements 
\fla{rbar}
\be
{\overline r}(\l,\e^Q Z^{-1}){\overline r}(\l\mu,\e^{P'}X)
{\overline r}(\mu,\e^{Q''}Z^{-1})=
{\overline r}(\mu,\e^P X){\overline r}(\l\mu,\e^{Q'}Z^{-1})
{\overline r}(\l,\e^{P''}X)
\label{qYBE}\ee
Here
\be 
\begin{array}{l}
P'=\exp\left({\widetilde H}(\l,\e^Q)\right)\circ P,\\
Q'=\exp\left({\widetilde H}(\mu,\e^P)\right)\circ Q,\\
P''=\exp\left({\widetilde H}(\l,\e^P)\right)\circ
\exp\left({\widetilde H}(\l\mu,\e^Q)\right)\circ P\\
Q''=\exp\left({\widetilde H}(\mu,\e^Q)\right)\circ
\exp\left({\widetilde H}(\l\mu,e^P)\right)\circ Q~.
\end{array}
\ee
where we have used the notation \fla{exp-action}.
Using \fla{PQbracket}
one can show that
\be\renewcommand{\arraystretch}{2.5}\begin{array}{ll}
\ds e^{NP'}
={1+\l^Ne^{NQ}\over \l^N+e^{NQ}}~e^{NP},&\ds
e^{NP''}
={1+\l^N\mu^Ne^{NQ'}\over \l^N\mu^N+e^{NQ'}}
~e^{NP},\\
\ds e^{NQ'}
={\mu^N+e^{NP}\over 1+\mu^Ne^{NP}}
~e^{NQ},&\ds
e^{NQ''}
={\l^N\mu^N+e^{NP'}\over 1+\l^N\mu^Ne^{NP'}}
~e^{NQ}~.
\end{array}\label{twistedpars}\renewcommand{\arraystretch}{1.5}\ee

The \YBE\ \fla{qYBE} contains four (complex) parameters:
two rapidities $\l$ and $\mu$ and two arbitrary parameters 
$P$ and $Q$. The connection of this equation with the star-triangle
 relation will be considered in Section~5.1. Note here two important 
properties of the $R$-matrix \fla{rbar}. Introducing 
\be
\w=1/\q_0^{2},\qquad \w^{1/2}=-1/\q_0, \label{omega}
\ee
where $\q_0$ is defined in \fla{define-roots} one can show
that
\be
{\overline r(\l,\w^n x)\over\overline r_l(x,z)}
=\left({1+\l^Nx^N\over 1+\l^{-N}x^N}\right)^{n\over N}
\prod^n_{j=1}
{1-\w^{-1/2}\l^{-1} x \w^j\over 1-\w^{-1/2}\l\phantom{{}^{-1}} x \w^j}
\label{prop-a}\ee
and 
\be
\prod^{N-1}_{j=0}\overline r(\l,\omega^jx)
=1~.\label{prop-b}
\ee

\nsection{Chiral Potts model.}

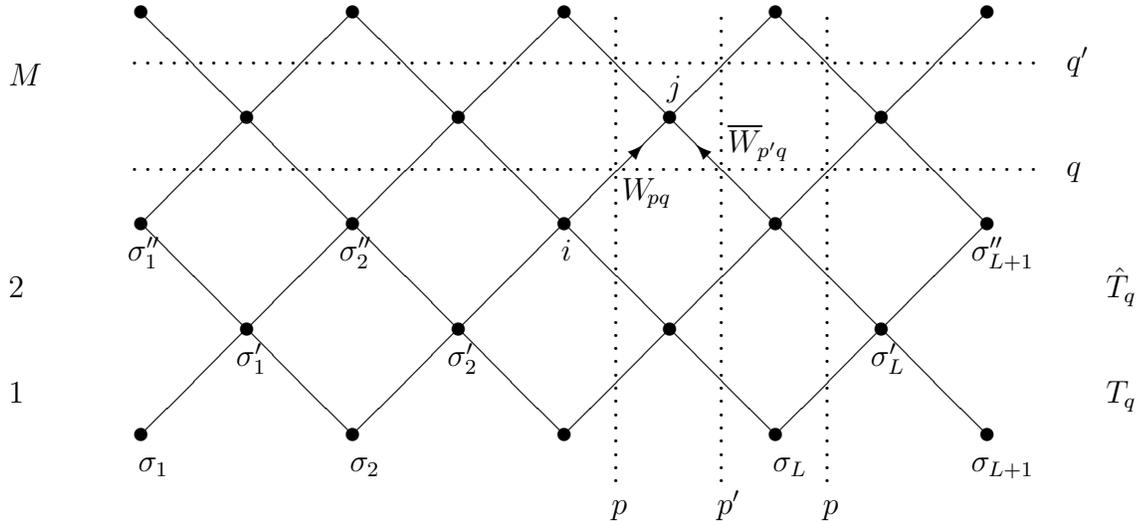
\begin{figure}[hbt]
\begin{picture}(420,200) (-40,0)
\multiput(0,0)(80,0){5}{\circle*{5}}
\multiput(0,80)(80,0){5}{\circle*{5}}
\multiput(0,160)(80,0){5}{\circle*{5}}
\multiput(40,40)(80,0){4}{\circle*{5}}
\multiput(40,120)(80,0){4}{\circle*{5}}
\multiput(-4,100)(5,0){69}{.}
\multiput(-4,140)(5,0){69}{.}
\multiput(178,-18)(0,5){36}{.}
\multiput(218,-18)(0,5){36}{.}
\multiput(258,-18)(0,5){36}{.}
\put(0,0) {\line (1,1) {160}}
\put(160,0) {\line (-1,1) {160}}
\put(160,0) {\line (1,1) {160}}
\put(320,0) {\line (-1,1) {160}}
\put(80,0) {\line (-1,1) {80}}
\put(0,80) {\line (1,1) {80}}
\put(80,0) {\line (1,1) {160}}
\put(240,0) {\line (-1,1) {160}}
\put(240,0) {\line (1,1) {80}}
\put(320,80) {\line (-1,1) {80}}
\thicklines
\put(184,104) {\vector (1,1) {6}}
\put(216,104) {\vector (-1,1) {6}}
\thinlines
\put(222,107) {$\overline{W}_{\! p'q}$}
\put(182,89) {$W_{\! pq}$}
\put(350,98)  {$q$}
\put(350,138) {$q'$}
\put(178,-30) {$p$}
\put(218,-30) {$p'$}
\put(258,-30) {$p$}
\put(-5,66) {$\sigma''_1$}
\put(75,66) {$\sigma''_2$} \put(-1,-14) {$\sigma_1$}
 \put(36,26) {$\sigma'_1$}
\put(79,-14) {$\sigma_2$} \put(116,26) {$\sigma'_2$}
 \put(239,-14){$\sigma_L$} \put(276,26) {$\sigma'_L$}
\put(314,-14) {$\sigma_{L+1}$}\put(314,66) {$\sigma''_{L+1}$}
\put(159,66) {$i$} \put(239,66) {} \put(199,128) {$j$} \put(365,12)
{$T_q$} \put(365,52) {$\hat{T}_q$} \put(-50,12) {1} \put(-50,52) {2}
\put(-50,132) {$M$} \end{picture}
\vspace{1.5cm}
 \caption{ The square lattice ${\cal L}$ of $M$
rows with $L$ sites per row. $T_q$ is the
transfer matrix of  an odd row, $\hat{T}_q$ of an even row.
Three vertical and two horizontal dotted rapidity lines
are shown.} \label{latticeL}
\end{figure}

  We define the chiral Potts model in the usual way \cite{BPA88},
\cite{BBP90}. Consider the square lattice $\cal L$, drawn diagonally
as in Fig.~\ref{latticeL}, with $L$ sites  per row. At
each site $i$ there is a spin $\sigma_i$, which takes values
$0, \ldots,  N-1$. There is an associated lattice $\cal L'$ denoted
by dotted lines, such that each edge of $\cal L$ passes through
a vertex of $\cal L'$. 
%\input{figure-cp}
%% The following picture is allowed a space 14cm by 6cm

To each vertical (horizontal) line on the ``dotted'' lattice $\cal L'$ assign
a rapidity variable $p$ ($q$). In general they may be different for
different lines. In fact a
convenient level of generality that we shall use here is to allow the
vertical rapidities to be alternating $p , p'
 , p , p' , \ldots $, as indicated\footnote{Here and below the letter
   $q$ denotes a rapidity variable; it is different from the parameter $\q$ 
in \fla{algebra}.}. Each edge of $\cal L$ is assigned 
a Boltzmann weight depending on the two spins adjacent to the edge
and on the two rapidities passing through the edge.
For example, consider a typical SW $\rightarrow$ NE
edge $(i,j)$  of $\cal L$ (with $j$ above $i$). 
The spins $\sigma_i, \sigma_j$ interact
 with Boltzmann weight $W_{pq}(\sigma_i - \sigma_j)$
(or $W_{p' q}(\sigma_i - \sigma_j)$ ). 
Similarly, on all SE $\rightarrow$ NW edges  the spins interact with Boltzmann
weight $\Wbar _{pq}(\sigma_i - \sigma_j)$ 
(or $\Wbar_{p' q}(\sigma_i - \sigma_j)$ ), where again $j$ is above
$i$. The explicit form of the Boltzmann weights will be given below.

Each rapidity variable $p$ (or $q$) is
represented by a four-vector $p=(a_p,b_p,c_p,d_p)$ which specify 
a point on the algebraic curve ${\cal C}_k$ defined by any two 
of the following four equations (the complimentary pair of 
equations follows from the other two)
\be
\begin{array}{cc}
a_p^N+k'b_p^N=kd_p^N, & k'a_p^N+b_p^N=kc_p^N, \\
ka_p^N+k'c_p^N=d_p^N, & kb_p^N+k'd_p^N=c_p^N, 
\end{array}\label{curve-abcd}
\ee
where $k^2+k'^2=1$. 
The modulus of the curve, $k$, is considered as a fixed parameter of
the model. It is convenient  to also use  another set of  
the ``$p$-variables'',  
\be 
\begin{array}{ccc}
x_p=a_p/d_p, & y_p=b_p/c_p, & s_p=d_p/c_p \label {def-xp}
\end{array}
\ee
and 
\be
t_p=x_p y_p = a_p b_p/c_p d_p. \label {def-tp}
\ee
In these variables the curve (\ref{curve-abcd}) reads 
\be\begin{array}{ccc}
x_p^N + y_p^N = k\,(1 + x_p^N y_p^N), & k x_p^N = 1 - k'
s_p^{-N}, &  k y_p^N = 1 - k' s_p^N  \label{curve-xy} 
\end{array}
\ee
With these definitions the Boltzmann weights have the form
\renewcommand{\arraystretch}{3}
\be\begin{array}{lll}
\displaystyle{\overline W_{pq}(n)\over \overline W_{pq}(0)}
&=
\displaystyle\prod^n_{j=1}{\omega a_pd_q-d_pa_q\omega^j\over
c_pb_q-b_pc_q\omega^j}
&\displaystyle=({s_p  s_q} )^n \prod_{j = 1}^{n}
{\omega x_p - \omega^j x_q\over y_q - \omega^j y_p },\\
\displaystyle{ W_{pq}(n)\over  W_{pq}(0)}
&\displaystyle=\prod^n_{j=1}{d_pb_q-a_pc_q\omega^j\over
b_pd_q-c_pa_q\omega^j}
&\displaystyle=({s_p /s_q })^n \prod_{j = 1}^{n}
{y_q - \omega^j x_p\over y_p - \omega^j x_q }
\end{array}
\label{def-W}
\ee
\renewcommand{\arraystretch}{1.5}
where $n \in {\Bbb Z}$ and $\omega$ is a primitive root of unity of
degree $N$.
The weights
satisfy the periodicity conditions $W_{pq}(n+N) = W_{pq} (n)$,
$\Wbar _{pq}(n+N) = \Wbar _{pq} (n)$. They also satisfy the star-triangle
relation \cite{BPA88}, \cite{AYP89}, \cite{MatSmirn90}:
\be 
\sum_{d = 0}^{N-1} \Wbar_{qr}(b-d) W_{pr}(a-d) \Wbar _{pq}(d-c) =
R_{pqr} W_{pq}(a-b) \Wbar_{pr} (b-c) W_{qr} (a-c).
\label{STR}
\ee
for all rapidities $p, q, r$ and all integers (spins) $a, b, c$. Here
$R_{pqr}$ is a spin-independent function, defined by 
\be
R_{pqr}=f_{pq} f_{qr} /f_{pr}, \label{def-Rpqr}
\ee
where
\be 
f_{pq}^N= \prod_{j=0}^{N-1}\left(\Wbar_{pq}^{(f)}(j)/W_{pq}(j)\right)
\ee
and 
\begin{equation}
\overline W^{(f)}_{pq}(n)
=\sum^{l-1}_{a=0}
\overline W_{pq}(a)\omega^{na}~.\label{def-Fourier}
\end{equation}
It follows from (\ref{def-W}) that 
\be 
\displaystyle{\overline W_{pq}^{(f)}(n)\over \overline W_{pq}^{(f)}(0)}
=
\displaystyle\prod^n_{j=1}{c_pb_q-a_pd_q\omega^j\over
b_pc_q-d_pa_q\omega^j}
=\prod_{j = 1}^{n}
{ y_q - \omega^j x_qs_ps_q\over y_p - \omega^j x_q s_ps_q }.\label{def-Wf}
\ee
Note also, that the normalization factor $\Wbar_{pq}^{(f)}(0)$ can
also be written in a product form by using the identity (2.44) of
\cite{BBP90}, namely
\be
 \prod_{j=0}^{N-1}\Wbar_{pq}^{(f)}(j)= \Wbar_{pq}(0)^N 
N^{N/2}  \, e^{i \pi
(N-1)(N-2)/12} \prod_{j=1}^{N-1} \frac{(t_p - \omega^j t_q )^j }
{ (x_p - \omega^j x_q )^j \; (y_p - \omega^j y_q )^j }.
\ee

We define row-to-row transfer matrices $T$ and $\hat{T}$ as in
\cite{BBP90}. Let $\sigma = \sigma_1 ,\ldots, \sigma_L$ be the spins
in the lower row
of  Figure~\ref{latticeL}. Similarly, let $\sigma ' =\sigma_1 ' ,\ldots,
\sigma_L '$
be the spins in
the next row, and $\sigma '' =\sigma_1 '' ,\ldots, \sigma_L ''$
those in the row above that.
Let $T$ be the $N^L$ by $N^L$ matrix with elements
\be \label{T}
T_{\sigma \sigma '} = \prod_{J=1}^{L} W_{pq}(\sigma_J - \sigma_J ')
 \Wbar_{p' q} (\sigma_{J+1} - \sigma_J ' ) ,\ee
similarly, let $\hat{T}$ be the  $N^L$ by $N^L$ matrix with elements
\be \label{That}
\hat{T}_{\sigma ' \sigma ''} = \prod_{J=1}^{L} \Wbar _{pq}(
\sigma_J '  - \sigma_J '')
 W_{p' q} (\sigma_{J} ' - \sigma_{J+1} '' ) \ee

Let $Y$ be a formal variable such that $Y^N=1$. Define
\be
\renewcommand{\arraystretch}{3}
\begin{array}{lll}
\displaystyle
F(p,q;Y)&\displaystyle
=N^{-1}\sum_{a=0}^{N-1}\sum_{b=0}^{N-1}\w^{-ab}W_{pq}(a)Y^b
&\displaystyle
=\sum_{a=0}^{N-1}W_{pq}^{(f)}(-a)Y^a\\
\displaystyle
\overline{F}(p,q;Y)&\displaystyle
=N^{-1}\sum_{a=0}^{N-1}\sum_{b=0}^{N-1}\w^{ab}
\Wbar_{pq}^{(f)}(a)Y^b
&\displaystyle
=\sum_{a=0}^{N-1}\Wbar_{pq}(a)Y^a
\end{array}\label{define-F}
\renewcommand{\arraystretch}{1.5}
\ee
where $W_{pq}^{(f)}(n)$ is defined similarly to (\ref{def-Fourier}).
It is easy to see that
\be
\begin{array}{cc}
F(p,q;\w^n)=W_{pq}(n), &\overline{F}(p,q;w^n)=\Wbar_{pq}^{(f)}(n)
\end{array}\label{diag-F}
\ee
for any $n\in {\Bbb Z}$ and that the functions $F$ and $\overline{F}$
are uniquely determined by the following recurrence relations
\be
{F(p,q;Y)\over F(p,q;\w^{-1}Y)}=
{s_p( y_q -  x_p Y)\over s_q(y_p -  x_q Y)},\qquad F(p,q;1)=W_{pq}(0)
\ee
\be
{\overline{F}(p,q;Y)\over\overline{F}(p,q;\w^{-1}Y)}=
{ y_q -  x_qs_ps_q Y\over y_p -  x_q s_ps_q Y }, \qquad
\overline{F}(p,q;1)= \Wbar_{pq}^{(f)}(0).
\label{recurrence-F}\ee

Now choose the following matrix realization of the algebra
\fla{XZ-alg} with $q_0^2=\w^{-1}$, $\w^N=1$, 
\be
(X)_{a,b}=\delta_{a,b+1}, \qquad (Z)_{a,b}=\w^a \delta_{a,b},
\ee
where $a,b\in {\Bbb Z_N}$ and  
\be
\delta_{a,b}=\left\{\begin{array}{ll}1,&a=b \pmod N,\\
                                        0,&\mbox{otherwise.}
                     \end{array}\right.
\ee
From \fla{diag-F} and \fla{define-F} it follows that 
\be
(F(p,q;Z))_{a,b}=\delta_{a,b} W_{pq}(a), \qquad
({\overline F}(p,q,X))_{a,b}=\sum_{k=0}^{N-1} {\overline W}_{pq}(k)
\delta_{a,b+k}.\label{mat-F}
\ee
Using these one  can  equivalently rewrite 
\cite{MatSmirn90} the \STR (\ref{STR}) as the
following relation between $F$ and $\overline F$ 
\be
F(p,q;Z^{-1})\overline F(p,r;X)F(q,r;Z^{-1})
=R^{-1}_{pqr}
\overline F(q,r;X)F(p,r;Z^{-1})\overline F(p,q;X)~.
\label{STR-a}\ee

\nsection{The connection of the discrete quantum Sine-Gordon and
chiral Potts models}

\subsection{The star-triangle relation}

We now want to identify equation \fla{qYBE} with the star-triangle
relation in the form \fla{STR-a}. Equation \fla{qYBE}
contains four continuous parameters $\l$, $\mu$, $P$, $Q$, while
\fla{STR-a} involves the modulus of the curve \fla{curve-abcd} and three
rapidities $p$, $q$, $r$, representing points on that
curve.
We have to establish a relationship between these two sets of
parameters. 
Denote
\be
\begin{array}{l}
A=1+(\e^Q/\l)^N,\qquad B=1+(\e^Q\l)^N\\
C=1+(\e^P/\mu)^N,\qquad D=1+(\e^P\mu)^N
\end{array}\label{A-D}
\ee
and consider the curve \fla{curve-abcd} with the modulus
\be
k^2=
{[AC(1-B)-BD(1-C)]
[BC(1-A)(1-D)-AB]\over
CD[A-B(1-D)][B-A-BC(1-A)]}
\label{module}
\ee
Now choose three rapidities variables $p$, $q$ and $r$
such that 
\be\begin{array}{ll}
\w^{-1/2}\l^{-1}\e^Q=x_p/y_q,\qquad &\w^{-1/2}\l\e^Q=x_q/y_p,\\
\w^{-1/2}\mu^{-1}\e^P=x_qs_qs_r/y_r,\qquad& 
\w^{-1/2}\mu\e^P=x_rs_rs_q/y_q,
\end{array}\label{first-four}
\ee
where $\w$ and $q_0$ are related by \fla{omega}. 
One can check that the last four equations are consistent with
\fla{module}. In fact, the expression \fla{module} is a corollary
of \fla{first-four}. This can be verified by direct substitution 
of \fla{first-four} into \fla{A-D} and \fla{module} with an account 
of the relations \fla{curve-xy}. Comparing now \fla{def-Wf} and the second 
equation in \fla{def-W} with equation \fla{prop-a} we see that
the relations \fla{first-four} allow us to identify the first factors 
in the left and right sides of \fla{STR-a} with the corresponding
factors in \fla{qYBE} 
\be
F(p,q,Z^{-1})={\overline r}(\l,e^{Q}Z^{-1}),\qquad 
\overline{F}(q,r,X)={\overline r}(\mu,e^{P}X),\label{fr-a}
\ee
provided the normalization factors $W_{pq}(0)$ and
$\overline{W}_{qr}(0)$ (which so far were at our disposal) are
appropriately chosen to satisfy \fla{prop-b}. To identify the
remaining factors in \fla{qYBE} and \fla{STR-a} we need 
four more pairs of relations
\be\begin{array}{ll}
\w^{-1/2}(\l\mu)^{-1}\e^{P'}=x_ps_ps_r/y_r,\qquad& 
\w^{-1/2}\l\mu\e^{P'}=x_rs_rs_p/y_p,\\
\w^{-1/2}(\l\mu)^{-1}\e^{Q'}=x_p/y_r,\qquad& 
\w^{-1/2}\l\mu\e^{Q'}=x_r/y_p,\\
\w^{-1/2}\l^{-1}\e^{P''}=x_ps_ps_q/y_q,\qquad& 
\w^{-1/2}\l\e^{P'}=x_rs_rs_p/y_p,\\
\w^{-1/2}\mu^{-1}\e^{Q''}=x_q/y_r,\qquad& \w^{-1/2}\mu\e^{Q''}=x_r/y_q.
\end{array}\label{last-eight}
\ee
Mysteriously enough, all these relation are corollaries 
of \fla{first-four}, \fla{twistedpars} and \fla{curve-xy}. The
calculations are simple but too long to be presented here. We quote
only a few formulae important for the derivation of \fla{last-eight}.
It follows from \fla{first-four} that
\be
\l^2=t_q/t_p, \qquad \mu^2=t_r/t_q,
\ee
where the $t$-variables are defined in \fla{def-tp} and that
\be
\e^{2Q}=\w x_px_q/y_py_q,\qquad \e^{2P}=\w x_qx_r s_q^2 s_r^2/y_qy_r.
\ee
Then one can rewrite \fla{twistedpars} as 
\be\begin{array}{ll}
\e^{2Q'}=\w x_px_r/y_py_r,\qquad &\e^{2P'}=\w x_px_r s_p^2 s_r^2/y_py_r,\\
\e^{2Q''}=\w x_px_r/y_py_r,\qquad &\e^{2P''}=\w x_px_q s_p^2 s_q^2/y_py_q.
\end{array}\ee
Similarly to \fla{fr-a} we have from \fla{last-eight} 
\be\begin{array}{ll}
\overline{F}(p,r,X)={\overline r}(\l\mu,e^{P'}X),\qquad 
&F(p,r,Z^{-1})={\overline r}(\l\mu,e^{Q'}Z^{-1}),\\ 
F(q,r,Z^{-1})={\overline r}(\mu,e^{Q''}Z^{-1}),\qquad 
&\overline{F}(p,q,X)={\overline r}(\l,e^{P''}X).\label{fr-b}
\end{array}
\ee
Thus we have proved the equivalence of the twisted \YBE\ \fla{qYBE}
with the  star-triangle relation of the chiral Potts model
\fla{STR-a}. Note that  the normalization of the weights 
of the chiral Potts model assumed by the relations \fla{fr-a}, \fla{fr-b}
is such that the factor $R_{pqr}$, \fla{def-Rpqr} in \fla{STR-a}
is equal to $1$, as it follows
from \fla{prop-b}.
 
Let us summarize the results of this section. The four relations  
\fla{first-four} allow us 
to express the parameters $k$, $p$, $q$, $r$ entering the \STR\ 
\fla{STR} or \fla{STR-a} through $\l$, $\mu$, $P$ and  $Q$ from the 
twisted \YBE\ \fla{qYBE} and vice versa. Remarkably, these four 
relations imply \fla{last-eight} enabling us to prove the
equivalence of the two \YBE s \fla{qYBE} and \fla{STR}.
This equivalence allows us to look at the chiral Potts model from a new 
angle.

\subsection{The evolution operator}
Consider now the evolution operator \fla{formof-U}. Substituting there the
expression for the $R$-matrix \fla{asymp} we can factorize $\bf U$
into classical and quantum parts \cite{BBR}
\be
{\bf U}={\bf U}_{\rm{cl}}{\bf U}_{\rm{quant}}
\ee
where
\be
{\bf U}_{\rm{cl}}=\exp\biggl(- {\cal H}_{\rm{cl}}^{(0)}/h\biggr)
\exp\biggl(-{\cal H}_{\rm{cl}}^{(1)}/h\biggr)\,(1+O(h)),
\ee
\be
{\cal H}_{\rm{cl}}^{(k)}=-\sum_{n=0}^{L-1}
{\widetilde H}(\kappa^2, w_{2n+k}), \qquad k=0,1,
\ee
with the function  ${\widetilde H}$ defined in \fla{Htilde} and
\fla{define-H}. Further,
\be
{\bf U}_{\rm{quant}}=\prod_{n=0}^{L-1}
{\overline r}(\kappa^2,\tau_{\rm{cl}}(w_{2n}))
\prod_{n=0}^{L-1}
{\overline r}(\kappa^2,w_{2n+1}),\label{Uquant}
\ee
\be
\tau_{\rm{cl}}(a)=\biggl(\exp({\cal H}_{\rm{cl}}^{(1)})\circ
\exp({\cal H}_{\rm{cl}}^{(0)})\circ a\biggr)=\lim_{h\rightarrow 0}
{\bf U}_{\rm{cl}}^{-1}\, a \,{\bf U}_{\rm{cl}}, 
\ee
where we used the notation \fla{adjoint-action}.
Let us now remove the constraints \fla{constraints} and consider 
a special case of the realization \fla{generators} 
when $q\rightarrow q_0$, $q_0^{2N}=1$,
\be
w_{2n}=\a_{2n}X_{n+1}, \qquad w_{2n+1}=\a_{2n+1}Z^{-1}_{n+1}Z_{n+2},
\qquad n=0,\ldots,L-1,\\ 
\ee
such that  
\be 
\a_{2n}=\a,\qquad \a_{2n+1}=\b, \qquad \forall n\ ,  \label{incond}
\ee
where $\a$ and $\b$ are fixed constants. 
It follows then from \fla{classical} that 
\be
\tau(\a_{2n})=\a,\qquad \tau(\a_{2n+1})=\b,\qquad  \forall n,
\ee
i.e., the configuration \fla{incond} is stationary with respect to 
the classical evolution. 

Consider the algebraic curve \fla{curve-abcd} with the modulus 
\be
k^2=\frac{(1-\a^N\b^N)\,(1-\a^N\b^{-N})}
{(1-\a^N\kappa^{2N})(1-\a^N\kappa^{-2N})}\ ,\label{new-ka}
\ee
and choose two rapidities $p$ and $q$ such that 
\be\begin{array}{ll}
\w^{-1/2}\b/\kappa^2=x_p/y_q,\qquad &\w^{-1/2}\b\kappa^2=x_q/y_p,\\
\w^{-1/2}\a/\kappa^2=x_ps_ps_q/y_q,\qquad& 
\w^{-1/2}\a\kappa^2=x_qs_ps_q/y_p,
\end{array}\label{twor}
\ee
Note that the last four equations are consistent with the relation 
\fla{new-ka}  and imply the latter as their corollary.
The expression \fla{Uquant} can now be rewritten as 
\be
{\bf U}_{\rm{quant}}=\prod_{n=1}^{L}
\overline{F}(p,q,X_{n})\  
\prod_{n=1}^{L}
{F}(p,q,Z^{-1}_{n}Z_{n+1}),\label{Uquant1}
\ee
where we have used \fla{define-F}. 
Taking into account \fla{mat-F} the matrix elements of this operator
are  
\be
\Big[{\bf U}_{\rm{quant}}\Big]_{a_1,a_2,\ldots,\a_{L}}^{b_1,b_2,\ldots,b_L}
=\prod_{n=1}^{L}
\overline{W}^{}_{pq}(a_n-b_n)\  
\prod_{n=1}^{L}
{W}_{pq}(b_{n+1}-b_n),\label{Uquant2}
\ee
where $W_{pq}(a-b)$ and $\overline{W}_{pq}(a-b)$ are the Boltzmann
weights of the chiral Potts model defined in \fla{def-W}.  

Let $M\ge1$ be an integer. Obviously, the trace 
\be
Z_{\mbox{chiral Potts}}=\mbox{Tr}\,\Big[{\bf U}_{\rm{quant}}\Big]^M
\ee
is the partition function of the chiral Potts model for a (non-diagonal) square
lattice of size $L$ by $M$ with periodic boundary conditions in
both directions (to avoid confusion, note that usually the periodic
boundary conditions in the chiral
Potts model are imposed 
for the diagonal (i.e, $45^\circ$-rotated) square lattice,
corresponding to the transfer matrix \fla{T}). 

\nsection{Concluding remarks}
We have presented a new interpretation of the chiral Potts model where it is
arising as a particular case of a more general Ising-type model
on a square lattice with local spins taking $N\ge2$ values at each
site. The Boltzmann weights of this model are determined by solutions 
of the classical discrete sine-Gordon model (which is an integrable
model of classical field theory). In this setting the chiral Potts
model corresponds to the simplest (constant) 
solution of the above classical model. It would be interesting to
consider more general spin models, corresponding to non-trivial
solutions of the ``background'' classical field theory. Our
construction also sheds some light on the origin of the ``non-difference''
property in the chiral Potts model. Further discussion of this point
will be given in \cite{BazCP}.

\nsection {Acknowledgements}
The author thanks R.J.Baxter, A.I.Bobenko, L.D.Faddeev, 
V.V.Mangazeev, S.M.Sergeev, Yu.A.Volkov and A.B.Zamolodchikov for
sharing their insights and useful discussions and 
N.Yu.Reshetikhin for collaboration on initial stages of this work. 
Special thanks to J.H.H.Perk for reading the manuscript and important 
remarks.

\app{Derivation of Eqs.\fla{comm}-\fla{result}.}

The derivation of the first equation in \fla{result} 
for $\Phi=\Phi_L$ goes through the 
following sequence of transformations by using \fla{ruv-equation} and 
\fla{uv-rel}
\be\begin{array}{l}
\Phi_L\,u =r(\lambda,u)r(\lambda\mu,v)r(\mu,u)\,u\\
=u\,r(\lambda,u)(\q\l\mu+v)(\q+\l\mu v)^{-1} r(\lambda\mu,v)r(\mu,u)\\
\displaystyle =u\,\left(\q\l\mu+v{1+q\l u\over \l+\q u}\right)
\left(\q+\l\mu v{\q\l+v\over \q+\l v}\right)^{-1} r(\lambda,u)
r(\lambda\mu,v)r(\mu,u)\\
=u g(u,v,\q)\Phi_L
\end{array}\label{trans-a}\ee
where the function $g$ is the same as in \fla{comm}. 

For $\Phi=\Phi_R$ the sequence of transformation is longer 
\be
\renewcommand{\arraystretch}{2.1}\begin{array}{l}
\Phi_R\,u=r(\mu,v)r(\lambda\mu,u)r(\lambda,v)\,u\\
\displaystyle =r(\mu,v)\,u\,r(\lambda\mu,u){\q\l+v\over \q+\l v}r(\lambda,v)\\
=r(\mu,v)\,u\, (\q\l (\l\mu +\q u) +v(1+\q\l\mu u))\times\\
\times
\biggl[(\q(\l\mu+\q u) +\l v(1+\q\l\mu u))^{-1}
(r(\lambda\mu,u)r(\lambda,v)\biggr]\\
=r(\mu,v) (\q\l (\l\mu +\q u) +\q^2 v(1+\q\l\mu u))\,u
\biggl[\ldots\biggr]\\
= (\q\l^2\mu +\q^2v+\q\l\,u(\q\mu+v))r(\mu,v)\,u
\biggl[\ldots\biggr]\\
=\biggl\{u\, (\q\l^2\mu +\q^2\l\mu u+v+\q\l v u)\displaystyle 
{\q\mu+v\over \q+\mu v}\biggr\}\,r(\mu,v)\,
\biggl[\ldots\biggr]\\
\displaystyle=\biggl\{ \ldots\biggr\}
\biggl(\l(\q\mu+v)+u{\q\mu+v\over \q+\mu v}
(\q^2 +\l^2 \mu \q^{-1} v)\biggr)^{-1}\Phi_R\\
=\displaystyle \biggl\{\ldots\biggr\}
{\q+\mu v\over \q\mu +v} (\q\l+ \q^2 u+\l\mu v+\q^2\l^2\mu v u)^{-1}
\Phi_R\\
=u g(u,v,\q)\Phi_R.
\end{array}\label{trans-b}
\renewcommand{\arraystretch}{1.5}
\ee
The second equation in \fla{result} is obtained from 
\fla{trans-a} and \fla{trans-b} with the help of the automorphims 
\fla{autho}.

%\bibliography{abbr,ref,total2}

\begin{thebibliography}{10}

\bibitem{Tsu86}
Tsuchiya, A. and Kanie, Y.
\newblock Vertex operators in conformal field theory on {${\bf P}\sp 1$} and
  monodromy representations of braid group.
\newblock In {\em Conformal field theory and solvable lattice models ({K}yoto,
  1986)}, volume~16 of {\em Adv. Stud. Pure Math.}, pages 297--372. Academic
  Press, Boston, MA, 1988.

\bibitem{Tsu89}
Tsuchiya, A., Ueno, K., and Yamada, Y.
\newblock Conformal field theory on universal family of stable curves with
  gauge symmetries.
\newblock In {\em Integrable systems in quantum field theory and statistical
  mechanics}, volume~19 of {\em Adv. Stud. Pure Math.}, pages 459--566.
  Academic Press, Boston, MA, 1989.

\bibitem{HKdN}
Howes, S., Kadanoff, L.~P., and den Nijs, M.
\newblock Quantum model for commensurate-incommensurate transitions.
\newblock Nucl. Phys.~B {\bf 215[FS7]} (1983) 169.

\bibitem{vG85}
von Gehlen, G. and Rittenberg, V.
\newblock $Z(n)$-symmetric quantum chains with an infinite set of conserved
  charges and $Z(n)$ zero modes.
\newblock Nucl. Phys. {\bf B257} (1985) 351.

\bibitem{AuY87}
Au-Yang, H., McCoy, B.~M., Perk, J. H.~H., Tang, S., and Yan, M.-L.
\newblock Commuting transfer matrices in the chiral Potts models: Solutions of
  star-triangle equations with genus $> 1$.
\newblock Phys. Lett. {\bf A123} (1987) 219--223.

\bibitem{BPA88}
Baxter, R.~J., Perk, J. H.~H., and Au-Yang, H.
\newblock New solutions of the star triangle relations for the chiral Potts
  model.
\newblock Phys. Lett. {\bf A128} (1988) 138--142.

\bibitem{CPH}
Baxter, R.~J.
\newblock Hyperelliptic Function Parameterization for the Chiral {P}otts Model.
\newblock In {\em Proceedings of the International Congress of Mathematicians},
  pages 1305--1317. Kyoto, Springer Verlag, 1990.

\bibitem{Kow}
Kowalewski, S.
\newblock Sur une propri\'et\'e du syst\`eme d'\'equations diff\'erentielles
  qui d\'efinit la rotation d'un corps solide autour d'un point fixe.
\newblock Acta Mathematica {\bf 12} (1889) 177--232.

\bibitem{Albertini:1988ux}
Albertini, G., McCoy, B.~M., Perk, J. H.~H., and Tang, S.
\newblock Excitation spectrum and order parameter for the integrable {N} state
  chiral {P}otts model.
\newblock Nucl. Phys. {\bf B314} (1989) 741.

\bibitem{Baxter:2005jt}
Baxter, R.~J.
\newblock Derivation of the order parameter of the chiral {P}otts model.
\newblock Phys. Rev. Lett. {\bf 94} (2005) 130602.

\bibitem{JMN93}
Jimbo, M., Miwa, T., and Nakayashiki, A.
\newblock Difference equations for the correlation functions of the
  eight-vertex model.
\newblock J. Phys. A {\bf 26} (1993) 2199--2209.

\bibitem{FZ82}
Fateev, V.~A. and Zamolodchikov, A.~B.
\newblock Self-dual solutions of the star-triangle relations in
  {$Z\sb{N}$}-models.
\newblock Phys. Lett. A {\bf 92} (1982) 37--39.

\bibitem{O49}
Onsager, L.
\newblock Discussion remark (Spontaneous magnetisation of the two-dimensional
  Ising model).
\newblock Nuovo Cimento (Suppl.) {\bf 6} (1949) 261.

\bibitem{Y52}
Yang, C.~N.
\newblock The spontaneous magnetization of a 2-dimensional Ising model.
\newblock Phys. Rev. {\bf 85} (1952) 808.

\bibitem{BS90}
Bazhanov, V.~V. and Stroganov, Y.~G.
\newblock Chiral {P}otts model as a descendant of the six-vertex model.
\newblock J. Statist. Phys. {\bf 59} (1990) 799--817.

\bibitem{Cherednik:1980ey}
Cherednik, I.~V.
\newblock On a method of constructing factorized {S} matrices in elementary
  functions.
\newblock Theor. Math. Phys. {\bf 43} (1980) 356--358.

\bibitem{BKMS}
Bazhanov, V.~V., Kashaev, R.~M., Mangazeev, V.~V., and Stroganov, Y.~G.
\newblock $({Z}\sb {N}\times)\sp {n-1}$ generalization of the chiral {P}otts
  model.
\newblock Comm. Math. Phys. {\bf 138} (1991) 393--408.

\bibitem{DCK90}
De~Concini, C. and Kac, V.
\newblock Representations of quantum groups at roots of $1$.
\newblock Progress in Math. {\bf 92} (1990) 471.

\bibitem{Date:1990bs}
Date, E., Jimbo, M., Miki, K., and Miwa, T.
\newblock Generalized chiral {P}otts models and minimal cyclic representations
  of ${U}_q(gl(n,{C}))$.
\newblock Commun. Math. Phys. {\bf 137} (1991) 133--148.

\bibitem{BB92}
Bazhanov, V. V.;~Baxter, R.~J.
\newblock New solvable lattice models in three dimensions.
\newblock J. Statist. Phys. {\bf 69} (1992) 453--485.

\bibitem{Z80}
Zamolodchikov, A.~B.
\newblock Tetrahedra equations and integrable systems in three-dimensional
  space.
\newblock Zh. Eksper. Teoret. Fiz. {\bf 79} (1980) 641--664.
\newblock [English trans.: Soviet Phys. JETP 52 (1980) ].

\bibitem{Z81}
Zamolodchikov, A.~B.
\newblock Tetrahedron equations and the relativistic $S$-matrix of
  straight-strings in $2+1$-dimensions.
\newblock Comm. Math. Phys. {\bf 79} (1981) 489--505.

\bibitem{B83}
Baxter, R.~J.
\newblock On {Z}amolodchikov's solution of the tetrahedron equations.
\newblock Comm. Math. Phys. {\bf 88} (1983) 185--205.

\bibitem{KMS93}
Kashaev, R.~M., Mangazeev, V.~V., and {Yu}.G., S.
\newblock Star-square and tetrahedron equations in the {B}axter-{B}azhanov
  model.
\newblock Int. J. Mod. Phys.~A {\bf 8} (1993) 1399--1409.

\bibitem{BB93}
Bazhanov, V.~V. and Baxter, R.~J.
\newblock Star-triangle relation for a three dimensional model.
\newblock J. Stat. Phys. {\bf 71} (1993) 839--864.

\bibitem{AYP89}
Au-Yang, H. and Perk, J.
\newblock Onsager's star-triangle equation: Master key to integrability.
\newblock In Aomoto, K. and Oda, T., editors, {\em Adv. Studies in Pure Math.},
  volume~19, pages 57--94. Academic/Kinokuniya, 1989.

\bibitem{FV:1993}
Faddeev, L. and Volkov, A.~Y.
\newblock Abelian current algebra and the Virasoro algebra on the lattice.
\newblock Phys. Lett. B {\bf 315} (1993) 311--318.

\bibitem{BKP}
Bobenko, A., Kutz, N., and Pinkall, U.
\newblock The discrete quantum pendulum.
\newblock Phys. Lett.~A {\bf 177} (1993) 399--404.

\bibitem{Faddeev:1994}
Faddeev, L.
\newblock Currentlike variables in massive and massless integrable models.
\newblock In {\em Quantum groups and their applications in physics (Varenna,
  1994)}, volume 127 of {\em Proc. Internat. School Phys. Enrico Fermi}, pages
  117--135, Amsterdam, 1996. IOS.

\bibitem{BBR}
Bazhanov, V.~V., Bobenko, A.~I., and Reshetikhin, N.~Y.
\newblock Quantum discrete {S}ine-{G}ordon model at roots of $1$: Integrable
  quantum system on the integrable classical background.
\newblock Commun. Math. Phys. {\bf 175} (1994) 377--400.

\bibitem{ZZ}
Zamolodchikov, A.~B. and Zamolodchikov, A.~B.
\newblock Factorized ${S}$-matrices in two dimensions as the exact solutions of
  certain relativistic quantum field theory models.
\newblock Ann. Physics {\bf 120} (1979) 253--291.

\bibitem{FT}
Faddeev, L.~D. and Takhtajan, L.~A.
\newblock {\em Hamiltonian methods in the theory of solitons}.
\newblock Springer, Berlin--New-York, 1987.

\bibitem{FKV:2001}
Faddeev, L.~D., Kashaev, R.~M., and Volkov, A.~Y.
\newblock Strongly coupled quantum discrete Liouville theory. I. Algebraic
  approach and duality.
\newblock Commun. Math. Phys. {\bf 219} (2001) 199--219.

\bibitem{Kharchev:2002}
Kharchev, S., Lebedev, D., and Semenov-Tian-Shansky, M.
\newblock Unitary representations of $U_{q}(\mathfrak{sl}(2,\mathbb{R}))$, the
  modular double, and the multiparticle q-deformed {T}oda chains.
\newblock Commun. Math. Phys. {\bf 225} (2002) 573--609.

\bibitem{Volkov:2005}
Volkov, A.~Y.
\newblock Noncommutative hypergeometry.
\newblock Comm. Math. Phys. {\bf 258} (2005) 257--273.

\bibitem{Bytsko:2006}
Bytsko, A.~G. and Teschner, J.
\newblock Quantization of models with non-compact quantum group symmetry:
  Modular XXZ magnet and lattice sinh-Gordon model.
\newblock J. Phys. {\bf A39} (2006) 12927--12981.

\bibitem{BMS07a}
Bazhanov, V.~V., Mangazeev, V.~V., and Sergeev, S.~M.
\newblock Faddeev-Volkov solution of the Yang-Baxter Equation and Discrete
  Conformal Symmetry.
\newblock Nucl. Phys. {\bf B784} (2007) 234--258.

\bibitem{BMS07b}
Bazhanov, V.~V., Mangazeev, V.~V., and Sergeev, S.~M.
\newblock Exact solution of the Faddeev-Volkov model.
\newblock Phys. Lett. A {\bf 372} (2008) 1547--1550.
\newblock arXiv.org:0706.3077.

\bibitem{BazCP}
Bazhanov, V.~V.
\newblock On star-triangle relation in the chiral Potts model.
\newblock in preparation, 2008.

\bibitem{Bazhanov:1995jpa}
Bazhanov, V.~V. and Reshetikhin, N.~Y.
\newblock Remarks on the quantum dilogarithm.
\newblock J. Phys. A {\bf 28} (1995) 2217--2226.

\bibitem{Bourbaki}
Bourbaki, N.
\newblock {\em Lie groups and Lie algebras}.
\newblock Elements of mathematics. Springer, Paris, 1972.

\bibitem{Berezin}
Berezin, F.~A.
\newblock {\em The method of second quantization}.
\newblock Academic Press, 1966.

\bibitem{FK}
Faddeev, L.~D. and Kashaev, R.~M.
\newblock Quantum dilogarithm.
\newblock Mod. Phys. Lett.~A {\bf 9} (1994) 427.

\bibitem{K}
Kirillov, A.~N.
\newblock Dilogarithm identities.
\newblock Preprint UTMS 94-56,, University of Tokyo, 1994.
\newblock hep-th/9408113.

\bibitem{BBP90}
Baxter, R.~J., Bazhanov, V.~V., and Perk, J. H.~H.
\newblock Functional relations for transfer matrices of the chiral {P}otts
  model.
\newblock Internat. J. Modern Phys. B {\bf 4} (1990) 803--870.

\bibitem{MatSmirn90}
Matveev, V.~B. and Smirnov, A.~O.
\newblock Some comments on the solvable chiral Potts model.
\newblock Lett. Math. Phys. {\bf 19} (1990) 179--185.

\end{thebibliography}
%\bibliographystyle{vvb-bibstyle}

\def\cprime{$'$} \def\cprime{$'$}

\end{document}